\documentclass[10pt]{article}
\usepackage{graphicx}
\usepackage{amsmath}
\usepackage{amssymb}
\usepackage{caption2}
\begin{document}
\begin{center}
{\large {\bf \sc{  Analysis of the  scalar, axialvector, vector, tensor doubly charmed tetraquark states  with QCD sum rules
  }}} \\[2mm]
Zhi-Gang  Wang \footnote{E-mail: zgwang@aliyun.com.  },  Ze-Hui Yan   \\
 Department of Physics, North China Electric Power University, Baoding 071003, P. R. China
\end{center}

\begin{abstract}
In this article, we construct the axialvector-diquark-axialvector-antidiquark type currents to interpolate the  scalar, axialvector, vector, tensor doubly charmed tetraquark states, and study them with QCD sum rules  systematically   by carrying out the operator product expansion up to the vacuum condensates of dimension 10 in a consistent way, the predicted masses can be confronted to the experimental data in the future. We can search for those doubly charmed tetraquark states in the  Okubo-Zweig-Iizuka  super-allowed strong decays to the charmed meson pairs.
\end{abstract}



\section{Introduction}
Recently, the LHCb collaboration observed the doubly charmed baryon  $\Xi_{cc}^{++}$ in the $\Lambda_c^+ K^- \pi^+\pi^+$ mass spectrum, and obtained the mass $M_{\Xi_{cc}^{++}}=3621.40 \pm 0.72 \pm 0.27 \pm 0.14\, \rm{MeV}$, but did not measure  the spin   \cite{LHCb-Xicc}.  The doubly heavy baryon configuration $QQq$ is very similar to the heavy-light
meson $\bar{Q}q$, where  we have a doubly heavy diquark $QQ$ instead of a heavy antiquark $\bar{Q}$ in color antitriplet. The attractive interaction induced by  one-gluon exchange  favors  formation  of  the diquarks in  color antitriplet \cite{One-gluon},  the favored configurations are the scalar ($C\gamma_5$) and axialvector ($C\gamma_\mu$) diquark states \cite{WangDiquark,WangLDiquark}. For the $cc$ quark system,  only the   axialvector  diquark $\varepsilon^{ijk} c^{T}_j C\gamma_\mu c_k$ and tensor diquark $\varepsilon^{ijk} c^{T}_j C\sigma_{\mu\nu }c_k$ survive due to the Fermi-Dirac  statistics,
the axialvector diquark $\varepsilon^{ijk} c^{T}_j C\gamma_{\mu} c_k$ is more stable than the tensor diquark $\varepsilon^{ijk} c^{T}_j C\sigma_{\mu\nu} c_k$, the observation   of the $\Xi_{cc}^{++}$ indicates that there exists strong  correlation between the two charm quarks.
We can take the diquark $\varepsilon^{ijk}  c^T_iC\gamma_\mu c_j$ as basic constituent to construct the spin $\frac{1}{2}$ current
\begin{eqnarray}
J_{\Xi_{cc}}(x)&=& \varepsilon^{ijk}  c^T_i(x)C\gamma_\mu c_j(x)\gamma_5\gamma^\mu u_k(x)  \, ,
\end{eqnarray}
or the spin $\frac{3}{2}$ current
\begin{eqnarray}
J_{\Xi_{cc}}^\mu(x)&=& \varepsilon^{ijk}  c^T_i(x)C\gamma^\mu c_j(x) u_k(x)  \, ,
\end{eqnarray}
to study the $\Xi_{cc}^{++}$ with the QCD sum rules \cite{Xicc-QCDSR}.

The doubly heavy tetraquark state $QQ\bar{q}\bar{q}^\prime$ is very similar to the doubly heavy baryon state $QQq$,
 where  we have a light antidiquark  $\bar{q}\bar{q}^\prime$ instead of a light quark $q$ in color triplet.
 The observation of the $\Xi_{cc}^{++}$ provides the crucial experimental input on the strong correlation between the two charm quarks, which may shed light on the
  spectroscopy of the doubly charmed  tetraquark states. An axialvector doubly charmed diquark state can combine with an axialvector or scalar  light antidiquark state to form a compact doubly charmed tetraquark state, it is interesting to revisit this subject with the QCD sum rules.
  The QCD sum rules is a powerful theoretical tool
in studying the ground state hadrons, and has  given many successful descriptions of
  the hadronic parameters on the phenomenological
side \cite{SVZ79,Reinders85}.
      Up to now, no experimental candidates for the doubly charmed tetraquark states  $cc\bar{q}\bar{q}^\prime$ or $qq^{\prime}\bar{c}\bar{c}$ have been observed. There have been several works on the doubly heavy tetraquark states, such as potential quark models \cite{QQ-quark-model,Karliner-Rosner}, QCD sum rules \cite{Nielsen-Lee,QQ-QCDSR,QQ-QCDSR-Chen}, heavy quark symmetry \cite{QQ-heavy-quark,Eichten-Quigg}, lattice QCD \cite{QQ-Latt,QQ-Latt-mass}, etc.

In previous work, we study the axialvector doubly heavy tetraquark states, which consist of  an axialvector diquark $\varepsilon^{ijk} Q^{T}_j C\gamma_{\mu} Q_k$ and a scalar antidiquark $\varepsilon^{ijk} \bar{q}^{T}_j \gamma_{5}C \bar{q}^{\prime}_k$, with the QCD sum rules in details by taking into account the energy scale dependence of the QCD spectral densities \cite{Wang-QQ-A}. In this article, we choose the axialvector diquark $\varepsilon^{ijk} c^{T}_j C\gamma_{\mu} c_k$ and axialvector antidiquark $\varepsilon^{ijk} \bar{q}^{T}_j \gamma_{\mu}C \bar{q}^{\prime}_k$ to construct the currents to interpolate the doubly charmed tetraquark states with the spin-parity $J^P=0^+,\,1^\pm,\,2^+$, and study them with the QCD sum rules systematically by taking into account the contributions of the vacuum condensates up to dimension 10 in a consistent way in the operator product expansion.

The article is arranged as follows:  we derive the QCD sum rules for the masses and pole residues of  the
  doubly charmed tetraquark states in Sect.2;  in Sect.3, we present the numerical results and discussions; and Sect.4 is reserved for our
conclusion.

\section{The QCD sum rules for  the   doubly charmed  tetraquark states }
In the following, we write down  the two-point correlation functions $\Pi_0(p)$, $\Pi_{\mu\nu\alpha\beta;1}(p)$ and $\Pi_{\mu\nu\alpha\beta;2}(p)$  in the QCD sum rules,
\begin{eqnarray}
\Pi_0(p)&=&i\int d^4x e^{ip \cdot x} \langle0|T\left\{J_0(x) J_0^{\dagger}(0)\right\}|0\rangle \, , \nonumber\\
\Pi_{\mu\nu\alpha\beta;1}(p)&=&i\int d^4x e^{ip \cdot x} \langle0|T\left\{J_{\mu\nu;1}(x) J_{\alpha\beta;1}^{\dagger}(0)\right\}|0\rangle \, , \nonumber\\
\Pi_{\mu\nu\alpha\beta;2}(p)&=&i\int d^4x e^{ip \cdot x} \langle0|T\left\{J_{\mu\nu;2}(x) J_{\alpha\beta;2}^{\dagger}(0)\right\}|0\rangle \, ,
\end{eqnarray}
where $J_0(x)=J_{\bar{u}\bar{d};0}(x),\,J_{\bar{u}\bar{s};0}(x),\,J_{\bar{s}\bar{s};0}(x)$, $J_{\mu\nu;1}(x)=J_{\mu\nu;\bar{u}\bar{d};1}(x),\,J_{\mu\nu;\bar{u}\bar{s};1}(x),\,J_{\mu\nu;\bar{s}\bar{s};1}(x)$,
$J_{\mu\nu;2}(x)=J_{\mu\nu;\bar{u}\bar{d};2}(x),\,J_{\mu\nu;\bar{u}\bar{s};2}(x),\,J_{\mu\nu;\bar{s}\bar{s};2}(x)$,
\begin{eqnarray}
J_{\bar{u}\bar{d};0}(x)&=&\varepsilon^{ijk}\varepsilon^{imn} \, c^{T}_j(x)C\gamma_\mu c_k(x) \,\bar{u}_m(x)\gamma^\mu C \bar{d}^T_n(x) \, , \nonumber\\
J_{\bar{u}\bar{s};0}(x)&=&\varepsilon^{ijk}\varepsilon^{imn} \, c^{T}_j(x)C\gamma_\mu c_k(x) \,\bar{u}_m(x)\gamma^\mu C \bar{s}^T_n(x) \, , \nonumber\\
J_{\bar{s}\bar{s};0}(x)&=&\varepsilon^{ijk}\varepsilon^{imn} \, c^{T}_j(x)C\gamma_\mu c_k(x) \,\bar{s}_m(x)\gamma^\mu C \bar{s}^T_n(x) \, , \nonumber\\
J_{\mu\nu;\bar{u}\bar{d};1}(x)&=&\varepsilon^{ijk}\varepsilon^{imn} \,\left[ c^{T}_j(x)C\gamma_\mu c_k(x) \,\bar{u}_m(x)\gamma_\nu C \bar{d}^T_n(x)-c^{T}_j(x)C\gamma_\nu c_k(x) \,\bar{u}_m(x)\gamma_\mu C \bar{d}^T_n(x) \right] \, , \nonumber\\
J_{\mu\nu;\bar{u}\bar{s};1}(x)&=&\varepsilon^{ijk}\varepsilon^{imn} \,\left[ c^{T}_j(x)C\gamma_\mu c_k(x) \,\bar{u}_m(x)\gamma_\nu C \bar{s}^T_n(x)-c^{T}_j(x)C\gamma_\nu c_k(x) \,\bar{u}_m(x)\gamma_\mu C \bar{s}^T_n(x) \right] \, , \nonumber\\
J_{\mu\nu;\bar{s}\bar{s};1}(x)&=&\varepsilon^{ijk}\varepsilon^{imn} \,\left[ c^{T}_j(x)C\gamma_\mu c_k(x) \,\bar{s}_m(x)\gamma_\nu C \bar{s}^T_n(x)-c^{T}_j(x)C\gamma_\nu c_k(x) \,\bar{s}_m(x)\gamma_\mu C \bar{s}^T_n(x) \right] \, , \nonumber\\
J_{\mu\nu;\bar{u}\bar{d};2}(x)&=&\varepsilon^{ijk}\varepsilon^{imn} \,\left[ c^{T}_j(x)C\gamma_\mu c_k(x) \,\bar{u}_m(x)\gamma_\nu C \bar{d}^T_n(x)+c^{T}_j(x)C\gamma_\nu c_k(x) \,\bar{u}_m(x)\gamma_\mu C \bar{d}^T_n(x) \right] \, , \nonumber\\
J_{\mu\nu;\bar{u}\bar{s};2}(x)&=&\varepsilon^{ijk}\varepsilon^{imn} \,\left[ c^{T}_j(x)C\gamma_\mu c_k(x) \,\bar{u}_m(x)\gamma_\nu C \bar{s}^T_n(x)+c^{T}_j(x)C\gamma_\nu c_k(x) \,\bar{u}_m(x)\gamma_\mu C \bar{s}^T_n(x) \right] \, , \nonumber\\
J_{\mu\nu;\bar{s}\bar{s};2}(x)&=&\varepsilon^{ijk}\varepsilon^{imn} \,\left[ c^{T}_j(x)C\gamma_\mu c_k(x) \,\bar{s}_m(x)\gamma_\nu C \bar{s}^T_n(x)+c^{T}_j(x)C\gamma_\nu c_k(x) \,\bar{s}_m(x)\gamma_\mu C \bar{s}^T_n(x) \right] \, , \nonumber\\
\end{eqnarray}
 the  $i$, $j$, $k$,  $m$, $n$ are color indexes, the $C$ is the charge conjugation matrix. We choose the currents $J_0(x)$, $J_{\mu\nu;1}(x)$ and $J_{\mu\nu;2}(x)$ to interpolate the spin-parity $J^P=0^+$, $1^\pm$ and $2^+$ doubly charmed tetraquark states, respectively.

On the phenomenological side,  we insert  a complete set of intermediate hadronic states with
the same quantum numbers as the current operators $J_0(x)$, $J_{\mu\nu;1}(x)$ and $J_{\mu\nu;2}(x)$ into the
correlation functions $\Pi_0(p)$, $\Pi_{\mu\nu\alpha\beta;1}(p)$ and $\Pi_{\mu\nu\alpha\beta;2}(p)$  respectively to obtain the hadronic representation
\cite{SVZ79,Reinders85}, and isolate the ground state
contributions,
\begin{eqnarray}
\Pi_{0}(p)&=&\frac{\lambda_{ Z}^2}{M_{Z}^2-p^2} +\cdots \, \, , \nonumber\\
&=&\Pi_{0}(p^2)\,\, , \\
\Pi_{\mu\nu\alpha\beta;1}(p)&=&\frac{\lambda_{ Z}^2}{M_{Z}^2-p^2}\left(p^2g_{\mu\alpha}g_{\nu\beta} -p^2g_{\mu\beta}g_{\nu\alpha} -g_{\mu\alpha}p_{\nu}p_{\beta}-g_{\nu\beta}p_{\mu}p_{\alpha}+g_{\mu\beta}p_{\nu}p_{\alpha}+g_{\nu\alpha}p_{\mu}p_{\beta}\right) \nonumber\\
&&+\frac{\lambda_{ Y}^2}{M_{Y}^2-p^2}\left( -g_{\mu\alpha}p_{\nu}p_{\beta}-g_{\nu\beta}p_{\mu}p_{\alpha}+g_{\mu\beta}p_{\nu}p_{\alpha}+g_{\nu\alpha}p_{\mu}p_{\beta}\right) +\cdots \, \, ,\\
&=&\Pi_Z(p^2)\left(p^2g_{\mu\alpha}g_{\nu\beta} -p^2g_{\mu\beta}g_{\nu\alpha} -g_{\mu\alpha}p_{\nu}p_{\beta}-g_{\nu\beta}p_{\mu}p_{\alpha}+g_{\mu\beta}p_{\nu}p_{\alpha}+g_{\nu\alpha}p_{\mu}p_{\beta}\right) \nonumber\\
&&+\Pi_Y(p^2)\left( -g_{\mu\alpha}p_{\nu}p_{\beta}-g_{\nu\beta}p_{\mu}p_{\alpha}+g_{\mu\beta}p_{\nu}p_{\alpha}+g_{\nu\alpha}p_{\mu}p_{\beta}\right) \, ,\nonumber\\
\Pi_{\mu\nu\alpha\beta;2} (p) &=&\frac{\lambda_{ Z}^2}{M_{Z}^2-p^2}\left( \frac{\widetilde{g}_{\mu\alpha}\widetilde{g}_{\nu\beta}+\widetilde{g}_{\mu\beta}\widetilde{g}_{\nu\alpha}}{2}-\frac{\widetilde{g}_{\mu\nu}\widetilde{g}_{\alpha\beta}}{3}\right) +\cdots \, \, , \nonumber\\
&=&\Pi_2(p^2)\left( \frac{\widetilde{g}_{\mu\alpha}\widetilde{g}_{\nu\beta}+\widetilde{g}_{\mu\beta}\widetilde{g}_{\nu\alpha}}{2}-\frac{\widetilde{g}_{\mu\nu}\widetilde{g}_{\alpha\beta}}{3}\right)\, , \end{eqnarray}
where $\widetilde{g}_{\mu\nu}=g_{\mu\nu}-\frac{p_{\mu}p_{\nu}}{p^2}$,  the  pole residues  $\lambda_{Z}$ and $\lambda_{Y}$ are defined by
\begin{eqnarray}
 \langle 0|J_0(0)|Z_{0^+}(p)\rangle &=& \lambda_{Z}     \, , \nonumber\\
  \langle 0|J_{\mu\nu;1}(0)|Z_{1^+}(p)\rangle &=& \frac{\lambda_{Z}}{M_Z} \, \epsilon_{\mu\nu\alpha\beta} \, \varepsilon^{\alpha}p^{\beta}\, , \nonumber\\
 \langle 0|J_{\mu\nu;1}(0)|Y_{1^-}(p)\rangle &=& \frac{\lambda_{Y}}{M_Y} \left(\varepsilon_{\mu}p_{\nu}-\varepsilon_{\nu}p_{\mu} \right)\, ,\nonumber\\
 \langle 0|J_{\mu\nu;2}(0)|Z_{2^+}(p)\rangle &=& \lambda_{Z} \, \varepsilon_{\mu\nu}   \, ,
\end{eqnarray}
the  $\varepsilon_\mu$ and $\varepsilon_{\mu\nu}$ are the polarization vectors of the spin $J=1$ and $2$ tetraquark states, respectively.
The summation of the polarization vectors $\varepsilon_\mu$ and $\varepsilon_{\mu\nu}$
 results in the following formula,
 \begin{eqnarray}
\sum_{\lambda}\varepsilon^*_{\mu}(\lambda,p)\varepsilon_{\nu}(\lambda,p)&=&-g_{\mu\nu}+\frac{p_\mu p_\nu}{p^2} \, ,\nonumber\\
\sum_{\lambda}\varepsilon^*_{\alpha\beta}(\lambda,p)\varepsilon_{\mu\nu}(\lambda,p)
 &=&\frac{\widetilde{g}_{\alpha\mu}\widetilde{g}_{\beta\nu}+\widetilde{g}_{\alpha\nu}\widetilde{g}_{\beta\mu}}{2}-\frac{\widetilde{g}_{\alpha\beta}\widetilde{g}_{\mu\nu}}{3}\,.
 \end{eqnarray}
The components $\Pi_{0}(p^2)$, $\Pi_{Z}(p^2)$, $\Pi_{Y}(p^2)$ and $\Pi_{2}(p^2)$ receive contributions of the hadronic states with the spin-parity $J^P=0^+$, $1^+$, $1^-$ and $2^+$, respectively.

Now we project out the components $\Pi_Z(p^2)$ and $\Pi_Y(p^2)$ by introducing the operators $P_Z^{\mu\nu\alpha\beta}$ and $P_Y^{\mu\nu\alpha\beta}$,
\begin{eqnarray}
\Pi_{1;A}(p^2)&=&p^2\Pi_Z(p^2)=P_Z^{\mu\nu\alpha\beta}\Pi_{\mu\nu\alpha\beta;1}(p) \, , \nonumber\\
\Pi_{1;V}(p^2)&=&p^2\Pi_Y(p^2)=P_Y^{\mu\nu\alpha\beta}\Pi_{\mu\nu\alpha\beta;1}(p) \, ,
\end{eqnarray}
where
\begin{eqnarray}
P_Z^{\mu\nu\alpha\beta}&=&\frac{1}{6}\left( g^{\mu\alpha}-\frac{p^\mu p^\alpha}{p^2}\right)\left( g^{\nu\beta}-\frac{p^\nu p^\beta}{p^2}\right)\, , \nonumber\\
P_Y^{\mu\nu\alpha\beta}&=&\frac{1}{6}\left( g^{\mu\alpha}-\frac{p^\mu p^\alpha}{p^2}\right)\left( g^{\nu\beta}-\frac{p^\nu p^\beta}{p^2}\right)-\frac{1}{6}g^{\mu\alpha}g^{\nu\beta}\, .
\end{eqnarray}

 In this article, we carry out the
operator product expansion for the correlation functions  $\Pi_0(p)$, $\Pi_{\mu\nu\alpha\beta;1}(p)$ and $\Pi_{\mu\nu\alpha\beta;2}(p)$ to the vacuum condensates  up to dimension-10, and take into account the vacuum condensates which are
vacuum expectations  of the operators  of the orders $\mathcal{O}( \alpha_s^{k})$ with $k\leq 1$ in a consistent way   \cite{WangHuang-3900,Wang-4660-2014,Wang-4025-CTP,WangHuang-NPA-2014,WangHuang-mole}, then we project out the components
  \begin{eqnarray}
\Pi_{1;A}(p^2)&=&P_Z^{\mu\nu\alpha\beta}\Pi_{\mu\nu\alpha\beta;1}(p) \, , \nonumber\\
\Pi_{1;V}(p^2)&=&P_Y^{\mu\nu\alpha\beta}\Pi_{\mu\nu\alpha\beta;1}(p) \, ,
\end{eqnarray}
on  the QCD side, and obtain the QCD spectral densities through dispersion relation,
\begin{eqnarray}
\rho_{0}(s)&=&\frac{{\rm Im}\Pi_{0}(s)}{\pi}\, , \nonumber\\
\rho_{1;A}(s)&=&\frac{{\rm Im}\Pi_{1;A}(s)}{\pi}\, , \nonumber\\
\rho_{1;V}(s)&=&\frac{{\rm Im}\Pi_{1;V}(s)}{\pi}\, , \nonumber\\
\rho_{2}(s)&=&\frac{{\rm Im}\Pi_{2}(s)}{\pi}\, ,
\end{eqnarray}
where $\rho_{0}(s)=\rho_{\bar{u}\bar{d};0}(s)$, $\rho_{\bar{u}\bar{s};0}(s)$, $\rho_{\bar{s}\bar{s};0}(s)$, $\rho_{1;A}(s)=\rho_{\bar{u}\bar{d};1;A}(s)$, $\rho_{\bar{u}\bar{s};1;A}(s)$, $\rho_{\bar{s}\bar{s};1;A}(s)$, $\rho_{1;V}(s)=\rho_{\bar{u}\bar{d};1;V}(s)$, $\rho_{\bar{u}\bar{s};1;V}(s)$, $\rho_{\bar{s}\bar{s};1;V}(s)$,  $\rho_{2}(s)=\rho_{\bar{u}\bar{d};2}(s)$, $\rho_{\bar{u}\bar{s};2}(s)$, $\rho_{\bar{s}\bar{s};2}(s)$. The explicit expressions of the QCD spectral densities are given in the Appendix.

Once the analytical expressions of the   QCD spectral densities $\rho_{0}(s)$,  $\rho_{1;A}(s)$, $\rho_{1;V}(s)$, $\rho_{2}(s)$ are obtained, we can  take the
quark-hadron duality below the continuum thresholds $s_0$ and perform Borel transform  with respect to
the variable $P^2=-p^2$ to obtain  the  QCD sum rules,
\begin{eqnarray}
\lambda^2_{Z/Y}\, \exp\left(-\frac{M^2_{Z/Y}}{T^2}\right)&=& \int_{4m_c^2}^{s_0} ds\, \rho(s) \, \exp\left(-\frac{s}{T^2}\right) \, ,
\end{eqnarray}
where $\rho(s)=\rho_{0}(s)$, $\rho_{1;A}(s)$, $\rho_{1;V}(s)$, $\rho_{2}(s)$.

We derive  Eq.(14) with respect to  $\tau=\frac{1}{T^2}$, then eliminate the
 pole residues   $\lambda_{Z/Y}$ to obtain the QCD sum rules for the masses of the doubly charmed tetraquark states,
 \begin{eqnarray}
 M^2_{Z/Y}&=& \frac{-\frac{d}{d \tau } \int_{4m_c^2}^{s_0} ds\,\rho(s)\,e^{-\tau s}}{\int_{4m_c^2}^{s_0} ds \,\rho(s)\,e^{-\tau s}}\, .
\end{eqnarray}

\section{Numerical results and discussions}

We take  the standard values of the vacuum condensates $\langle
\bar{q}q \rangle=-(0.24\pm 0.01\, \rm{GeV})^3$,   $\langle
\bar{q}g_s\sigma G q \rangle=m_0^2\langle \bar{q}q \rangle$,
$m_0^2=(0.8 \pm 0.1)\,\rm{GeV}^2$, $\langle\bar{s}s \rangle=(0.8\pm0.1)\langle\bar{q}q \rangle$, $\langle\bar{s}g_s\sigma G s \rangle=m_0^2\langle \bar{s}s \rangle$,  $\langle \frac{\alpha_s
GG}{\pi}\rangle=(0.33\,\rm{GeV})^4 $    at the energy scale  $\mu=1\, \rm{GeV}$
\cite{SVZ79,Reinders85,Colangelo-Review}, and choose the $\overline{MS}$ masses $m_{c}(m_c)=(1.275\pm0.025)\,\rm{GeV}$, $m_s(\mu=2\,\rm{GeV})=(0.095\pm0.005)\,\rm{GeV}$ from the Particle Data Group \cite{PDG}.
Moreover, we take into account the energy-scale dependence of  the input parameters on the QCD side,
\begin{eqnarray}
\langle\bar{q}q \rangle(\mu)&=&\langle\bar{q}q \rangle(Q)\left[\frac{\alpha_{s}(Q)}{\alpha_{s}(\mu)}\right]^{\frac{4}{9}}\, , \nonumber\\
 \langle\bar{s}s \rangle(\mu)&=&\langle\bar{s}s \rangle(Q)\left[\frac{\alpha_{s}(Q)}{\alpha_{s}(\mu)}\right]^{\frac{4}{9}}\, , \nonumber\\
 \langle\bar{q}g_s \sigma Gq \rangle(\mu)&=&\langle\bar{q}g_s \sigma Gq \rangle(Q)\left[\frac{\alpha_{s}(Q)}{\alpha_{s}(\mu)}\right]^{\frac{2}{27}}\, , \nonumber\\ \langle\bar{s}g_s \sigma Gs \rangle(\mu)&=&\langle\bar{s}g_s \sigma Gs \rangle(Q)\left[\frac{\alpha_{s}(Q)}{\alpha_{s}(\mu)}\right]^{\frac{2}{27}}\, , \nonumber\\
m_c(\mu)&=&m_c(m_c)\left[\frac{\alpha_{s}(\mu)}{\alpha_{s}(m_c)}\right]^{\frac{12}{25}} \, ,\nonumber\\
m_s(\mu)&=&m_s({\rm 2GeV} )\left[\frac{\alpha_{s}(\mu)}{\alpha_{s}({\rm 2GeV})}\right]^{\frac{4}{9}} \, ,\nonumber\\
\alpha_s(\mu)&=&\frac{1}{b_0t}\left[1-\frac{b_1}{b_0^2}\frac{\log t}{t} +\frac{b_1^2(\log^2{t}-\log{t}-1)+b_0b_2}{b_0^4t^2}\right]\, ,
\end{eqnarray}
  where $t=\log \frac{\mu^2}{\Lambda^2}$, $b_0=\frac{33-2n_f}{12\pi}$, $b_1=\frac{153-19n_f}{24\pi^2}$, $b_2=\frac{2857-\frac{5033}{9}n_f+\frac{325}{27}n_f^2}{128\pi^3}$,  $\Lambda=213\,\rm{MeV}$, $296\,\rm{MeV}$  and  $339\,\rm{MeV}$ for the flavors  $n_f=5$, $4$ and $3$, respectively  \cite{PDG}, and evolve all the input parameters to the optimal energy scales   $\mu$ to extract the masses of the doubly charmed tetraquark states $Z$ and $Y$.

In the article, we study the doubly charmed tetraquark states, the two charm quarks form an axialvector  doubly charmed  diquark state in color antitriplet, the axialvector doubly charmed  diquark state serves as a static well potential and combines with an axialvector  light  antidiquark state in color triplet to form a compact tetraquark state. While in the hidden-charm tetraquark states, the charm quark $c$ serves as a static well potential and  combines with the light quark $q$  to form a charmed diquark  in  color antitriplet,
 the charm antiquark $\bar{c}$ serves  as another static well potential and combines with the light antiquark $\bar{q}^\prime$  to form a charmed antidiquark in  color triplet, then the charmed diquark and charmed antidiquark combine together to form a hidden-charm tetraquark state. The quark structures of the doubly charmed tetraquark states and hidden-charm tetraquark states are quite different.

In Refs.\cite{WangHuang-3900,Wang-4660-2014,Wang-4025-CTP,WangHuang-NPA-2014,WangHuang-mole}, we study the acceptable energy scales of the QCD spectral densities  for the hidden-charm (hidden-bottom) tetraquark states and molecular states   in the QCD sum rules in details  for the first time,  and suggest an energy scale formula  $\mu=\sqrt{M^2_{X/Y/Z}-(2{\mathbb{M}}_Q)^2}$ to determine  the optimal   energy scales.
The  energy scale formula also works well in studying the hidden-charm pentaquark states \cite{WangPc}. The updated values of the effective heavy quark masses are ${\mathbb{M}}_c=1.82\,\rm{GeV}$ and ${\mathbb{M}}_b=5.17\,\rm{GeV}$ \cite{Wang-1601}. It is not necessary  for the effective charm quark mass ${\mathbb{M}}_c$ in the doubly charmed tetraquark states to have the same value as the one in the hidden-charm tetraquark states. In calculations, we observe that if we choose a slightly different value ${\mathbb{M}}_c=1.84\,\rm{GeV}$, the criteria of the QCD sum rules can be satisfied more easily.   We obtain the energy scale formula  by setting the energy scale $\mu=V$, the  virtuality $V$  (or bound energy not as robust) is defined by $V=\sqrt{M^2_{X/Y/Z}-(2{\mathbb{M}}_c)^2}$ \cite{Wang-4660-2014,Wang-4025-CTP}. In this article, we take into account the $SU(3)$ breaking effect $m_s(\mu)$ by subtracting the $m_s(\mu)$ from the  virtuality $V$, $\mu_k=V_k=\sqrt{M^2_{X/Y/Z}-(2{\mathbb{M}}_c)^2}-k\,m_s(\mu_k)$, where the numbers of the strange antiquark $\bar{s}$ in the doubly charmed tetraquark states are $k=0,1,2$.

In this article, we take the continuum threshold parameters  as  $\sqrt{s_0}=M_{Z/Y}+(0.4\sim0.7)\,\rm{GeV}$, and vary the parameters $\sqrt{s_0}$ to obtain   the optimal Borel parameters $T^2$ to satisfy  the  following four criteria:

$\bf 1.$ Pole dominance on the phenomenological side;

$\bf 2.$ Convergence of the operator product expansion;

$\bf 3.$ Appearance of the Borel platforms;

$\bf 4.$ Satisfying the energy scale formula.

The resulting  Borel parameters or Borel windows $T^2$, continuum threshold parameters $s_0$, optimal energy scales of the QCD spectral densities, pole contributions of the ground states are shown explicitly in Table 1. From Table 1, we can see that the pole dominance can be well satisfied.
The pole contributions $\rm{PC}$ are defined by
 \begin{eqnarray}
{\rm PC}&=& \frac{  \int_{4m_c^2}^{s_0} ds\,\rho(s)\,\exp\left(-\frac{s}{T^2}\right)}{\int_{4m_c^2}^{\infty} ds \,\rho(s)\,\exp\left(-\frac{s}{T^2}\right)}\, ,
\end{eqnarray}
which decrease  monotonously and quickly  with increase of the Borel parameter $T^2$, as the continuum contributions are depressed by the factor  $\exp\left(-\frac{s}{T^2}\right)$, large Borel parameter $T^2$ enhances the continuum contributions, the largest power of the QCD spectral densities $\rho(s)\propto s^4$, the convergent behaviors of the operator product expansion are  not very good for the tetraquark states and molecular states. Furthermore, the pole contributions increase  monotonously   with increase of the threshold parameters $s_0$, the uncertainties of the threshold parameters $\delta \sqrt{s_0}=\pm0.1\,\rm{GeV}$  also lead to rather large variations  of the pole contributions.
So in the small Borel window $T^2_{max}-T^2_{min}=0.4\,\rm{GeV}^2$ for the $J^P=(0/1/2)^+$ tetraquark states, the pole contributions vary in a rather large range, about $(40-60)\%$. Although the pole contributions have rather large uncertainties,  ${\rm PC}=(50\pm 10)\%$ for the $J^P=(0/1/2)^+$ tetraquark states and ${\rm PC}=(60\pm 10)\%$ for the $J^P=1^-$ tetraquark states, the  pole dominance can be well satisfied, the predictions are reliable.
On the other hand, if we choose larger energy scales $\mu$, the pole contributions are enhanced, the pole contributions are less sensitive to the Borel parameter $T^2$, however, we should determine the energy scales of the QCD spectral densities in a consistent way by using the energy scale formula.

   In Fig.1, we plot the absolute  contributions of the vacuum condensates $|D(n)|$ in the operator product expansion  for the central values of the input parameters,
   \begin{eqnarray}
D(n)&=& \frac{  \int_{4m_c^2}^{s_0} ds\,\rho_{n}(s)\,\exp\left(-\frac{s}{T^2}\right)}{\int_{4m_c^2}^{s_0} ds \,\rho(s)\,\exp\left(-\frac{s}{T^2}\right)}\, ,
\end{eqnarray}
where the $\rho_{n}(s)$ are the QCD spectral densities for the vacuum condensates of dimension $n$.
    From the figure, we can see that the dominant contributions come from the perturbative terms (or $D(0)$) for the $1^-$ tetraquark states, the operator product expansion is well convergent, while in the case of the $0^+$, $1^+$ and $2^+$ tetraquark states, the contributions of the vacuum condensates of dimension $n=6$ are very large, but the contributions of the vacuum condensates of dimensions $6, \,8,\,10$ have the hierarchy $|D(6)|\gg |D(8)|\gg |D(10)|$, the operator product expansion is also convergent.

\begin{figure}
 \centering
 \includegraphics[totalheight=5cm,width=7cm]{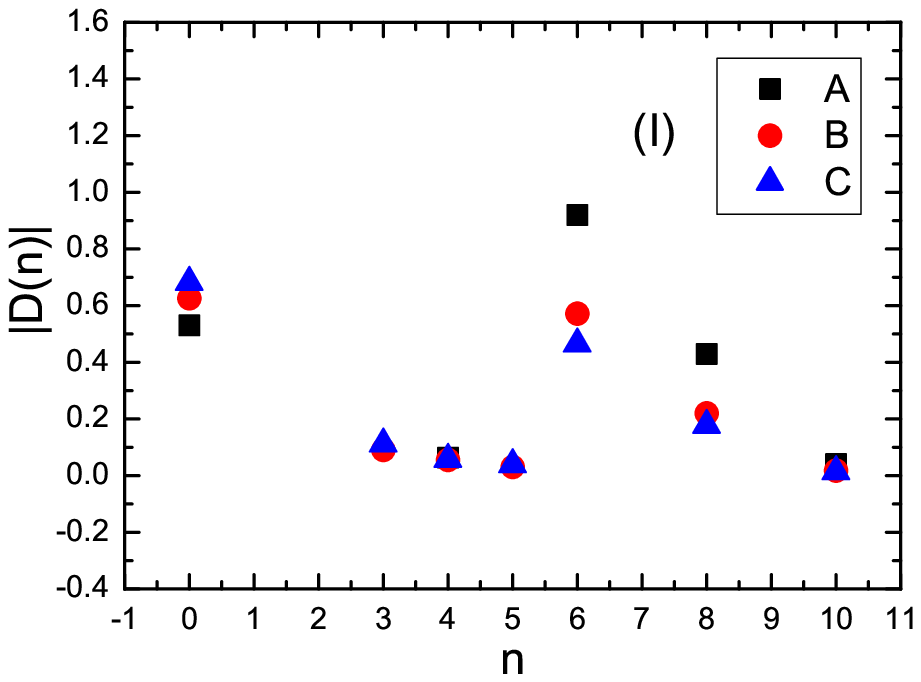}
 \includegraphics[totalheight=5cm,width=7cm]{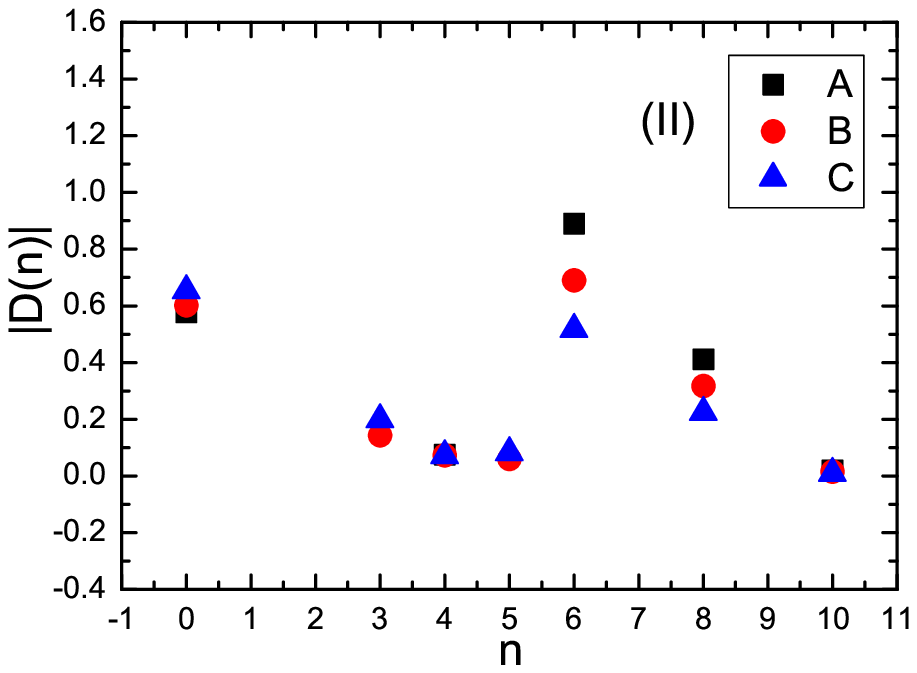}
 \includegraphics[totalheight=5cm,width=7cm]{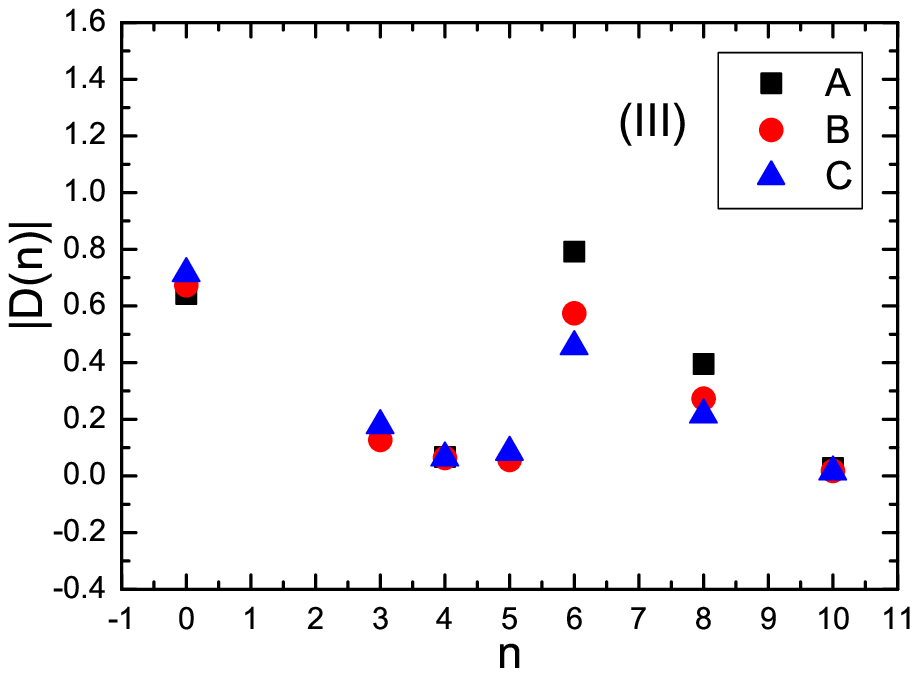}
 \includegraphics[totalheight=5cm,width=7cm]{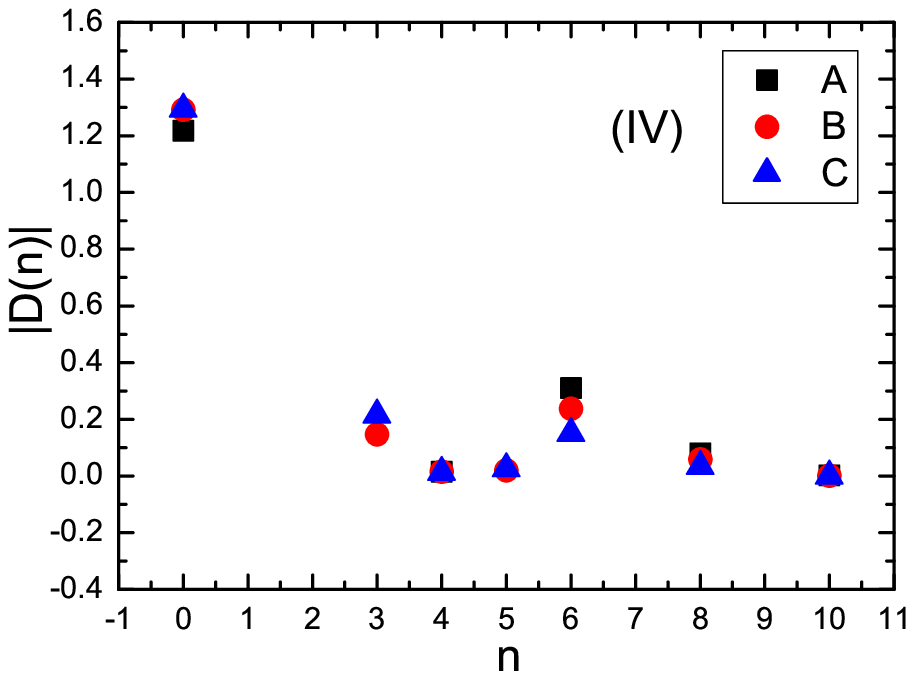}
         \caption{ The absolute contributions  of the vacuum condensates of dimension $n$ for central values of the input parameters, where  the (I), (II), (III) and  (IV) denote the tetraquark states with $J^P=0^+$, $1^+$, $2^+$ and $1^-$ respectively, the $A$, $B$ and $C$ denote the quark constituents $cc\bar{u}\bar{d}$, $cc\bar{u}\bar{s}$ and $cc\bar{s}\bar{s}$ respectively.   }
\end{figure}

We take  into account all uncertainties of the input parameters,
and obtain the values of the masses and pole residues of
 the   doubly charmed tetraquark states $Z$ and $Y$, which are  shown explicitly in Table 1 and Figs.2-5. In Figs.2-5, we plot the masses and pole residues of the doubly charmed tetraquark states in  much large ranges than the Borel windows.
From Figs.2-5, we can see that there appear platforms in  the Borel windows shown in Table 1.  Furthermore, from Table 1, we can see that the energy scale formula $\mu_k=\sqrt{M^2_{X/Y/Z}-(2{\mathbb{M}}_c)^2}-k\,m_s(\mu_k)$ with $k=0,1,2$ is also satisfied. Now the four criteria are all satisfied,  we expect to make reliable predictions.

\begin{figure}
 \centering
 \includegraphics[totalheight=5cm,width=7cm]{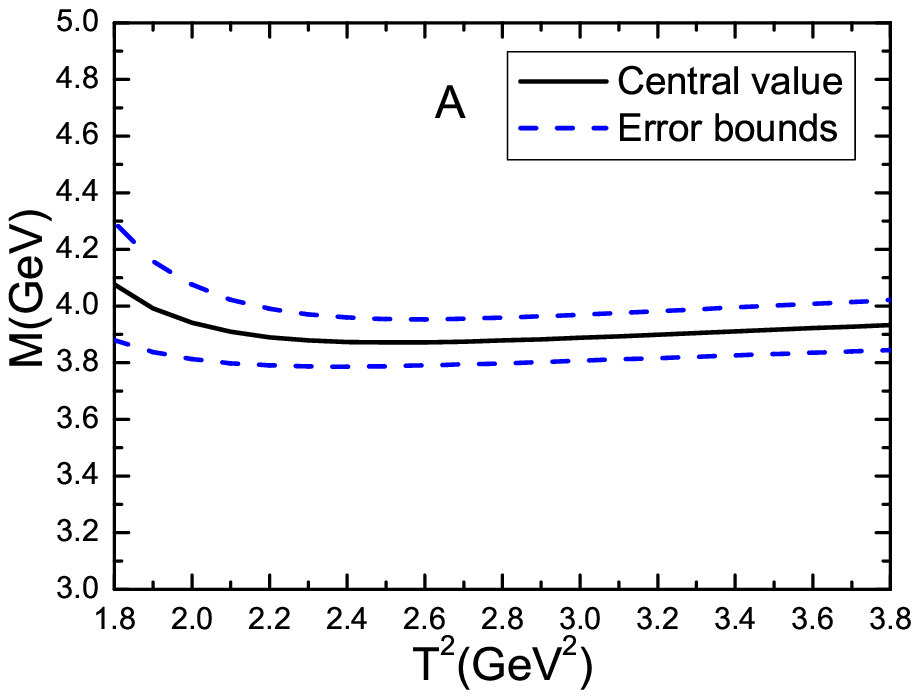}
 \includegraphics[totalheight=5cm,width=7cm]{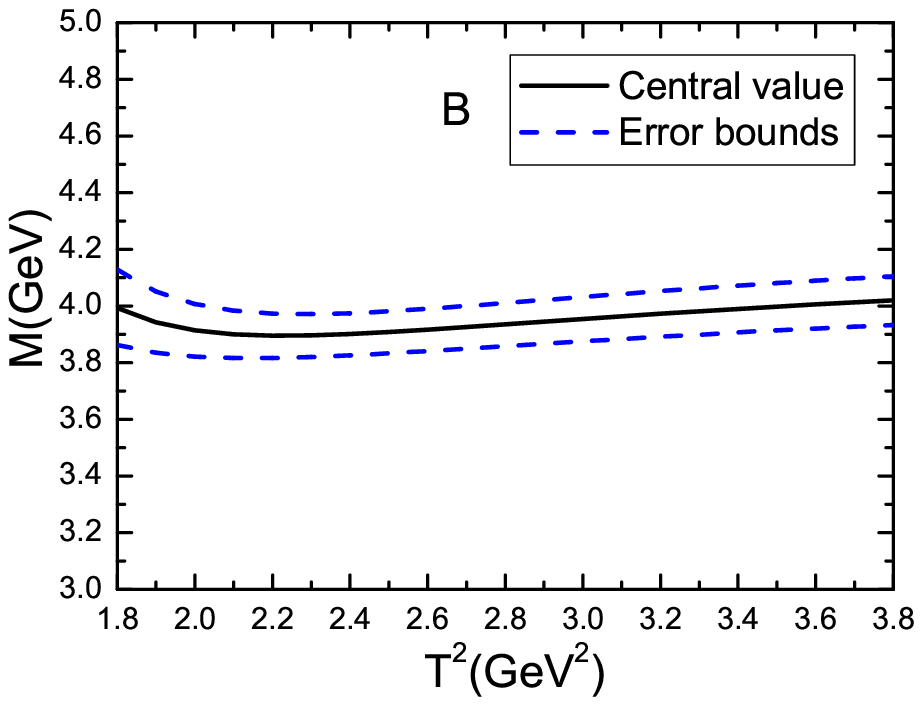}
 \includegraphics[totalheight=5cm,width=7cm]{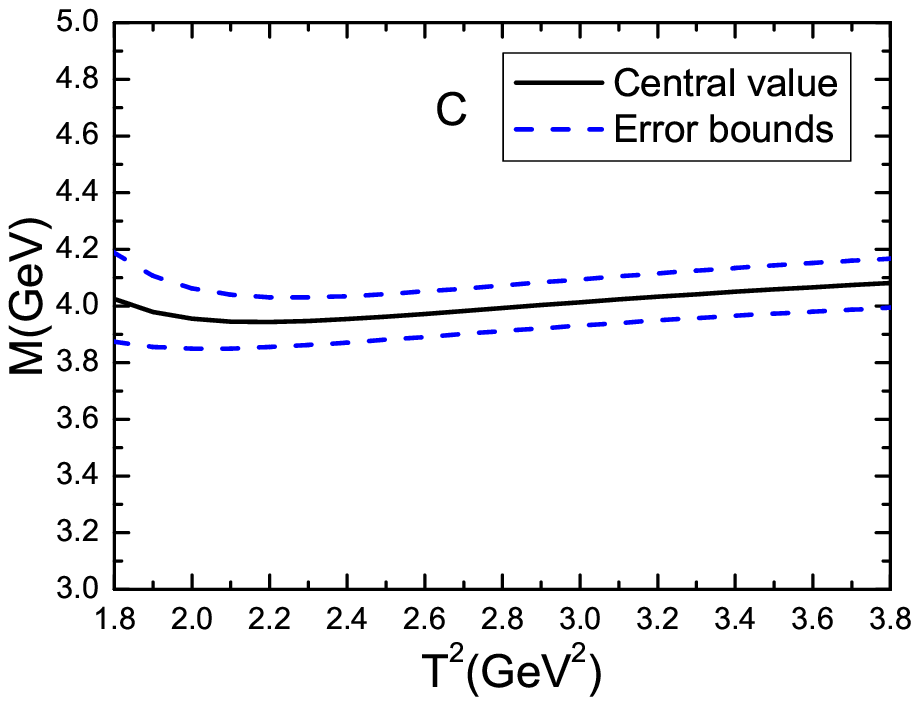}
 \includegraphics[totalheight=5cm,width=7cm]{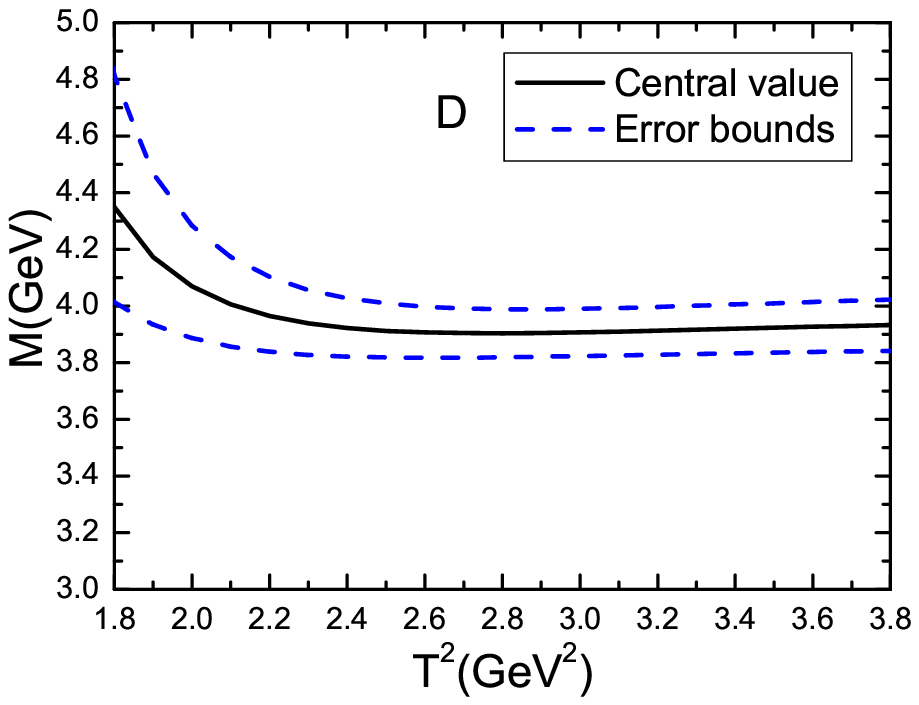}
 \includegraphics[totalheight=5cm,width=7cm]{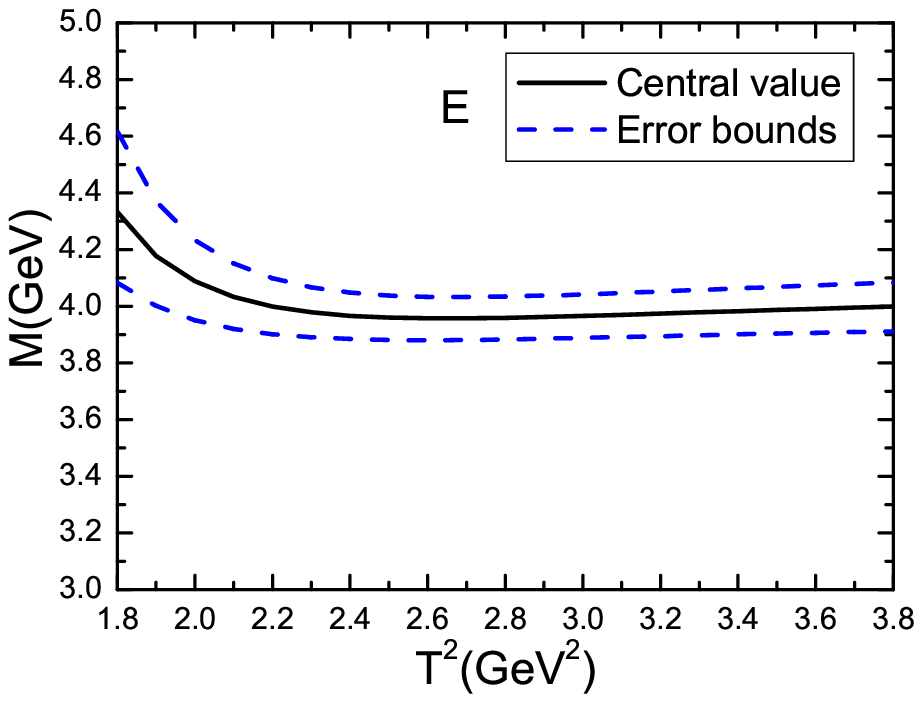}
 \includegraphics[totalheight=5cm,width=7cm]{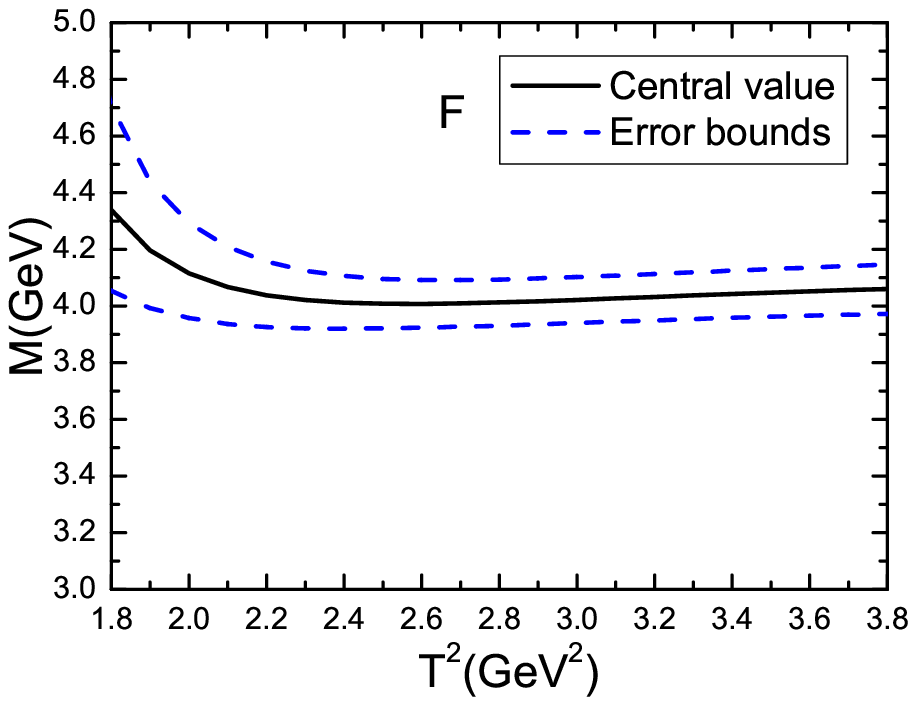}
         \caption{ The masses with variations of the Borel parameters, where  the $A$, $B$, $C$, $D$, $E$ and $F$ denote the tetraquark states  $cc\bar{u}\bar{d}\,(0^+)$, $cc\bar{u}\bar{s}\,(0^+)$, $cc\bar{s}\bar{s}\,(0^+)$, $cc\bar{u}\bar{d}\,(1^+)$, $cc\bar{u}\bar{s}\,(1^+)$ and $cc\bar{s}\bar{s}\,(1^+)$, respectively.   }
\end{figure}

\begin{figure}
 \centering
 \includegraphics[totalheight=5cm,width=7cm]{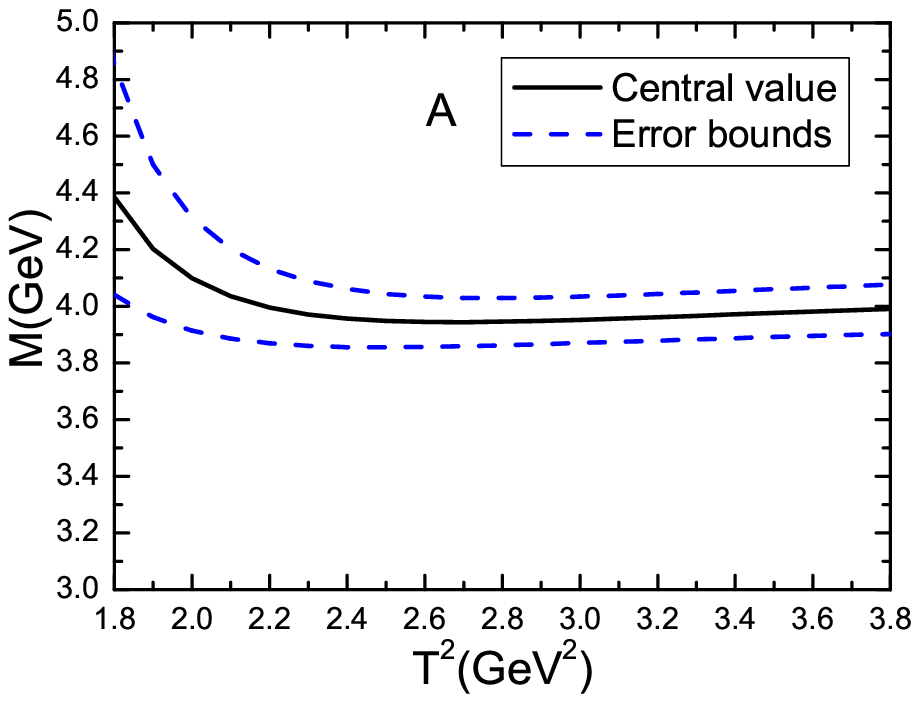}
 \includegraphics[totalheight=5cm,width=7cm]{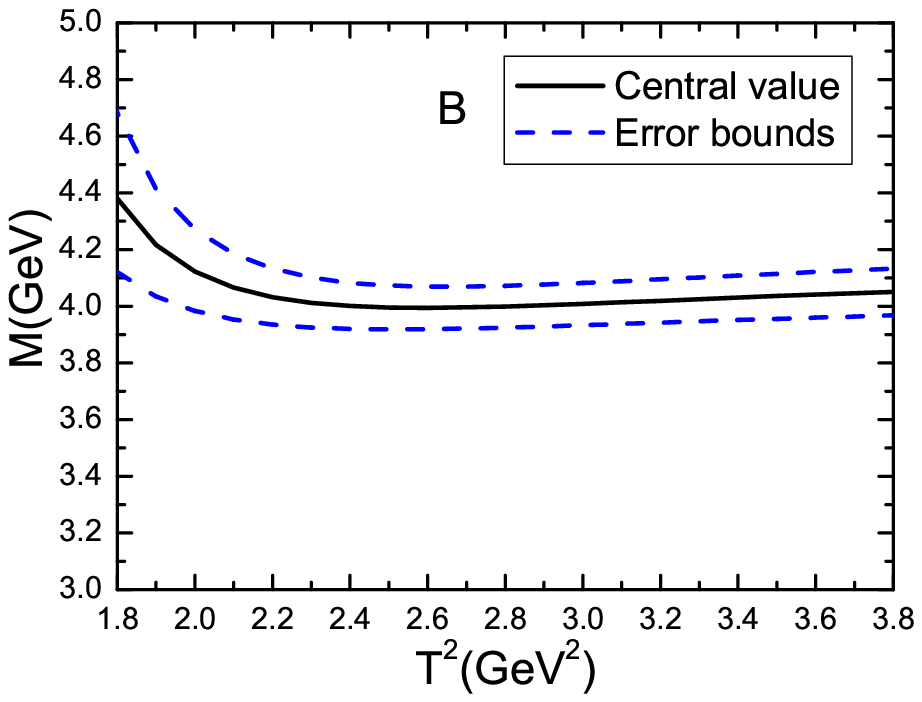}
 \includegraphics[totalheight=5cm,width=7cm]{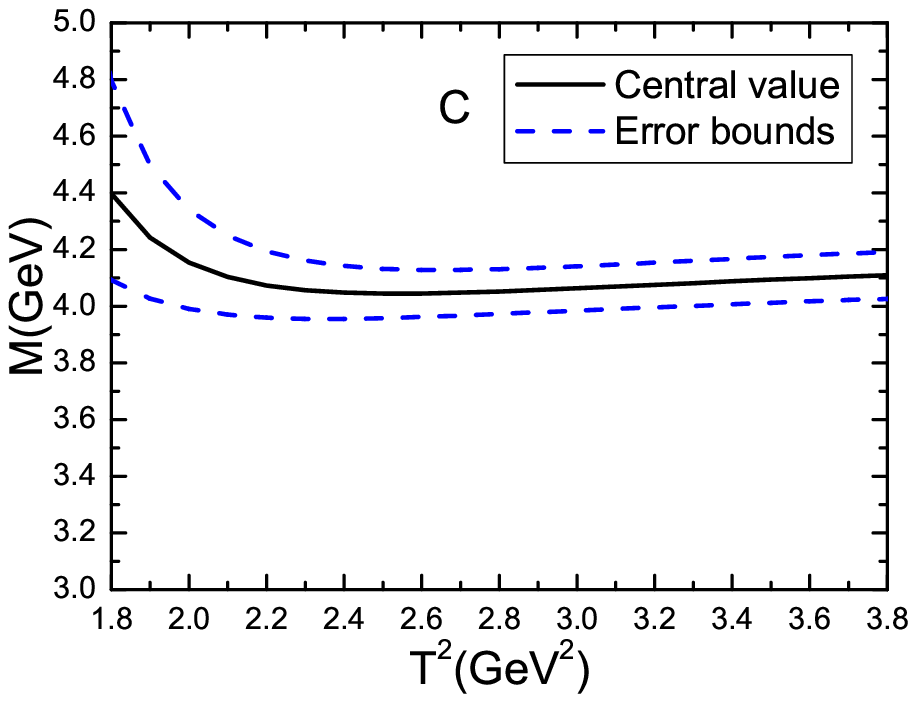}
 \includegraphics[totalheight=5cm,width=7cm]{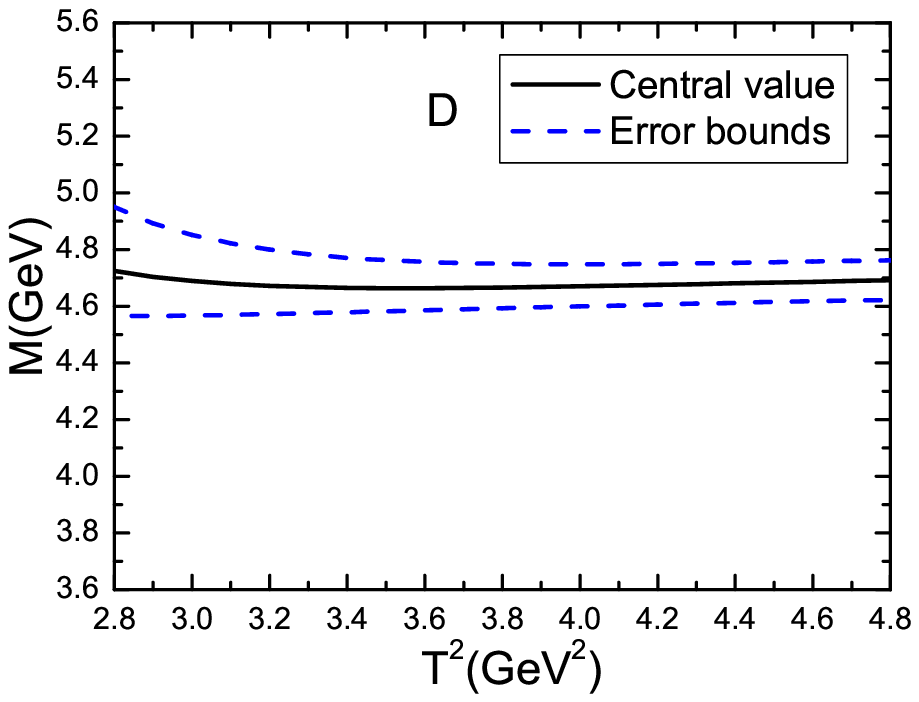}
 \includegraphics[totalheight=5cm,width=7cm]{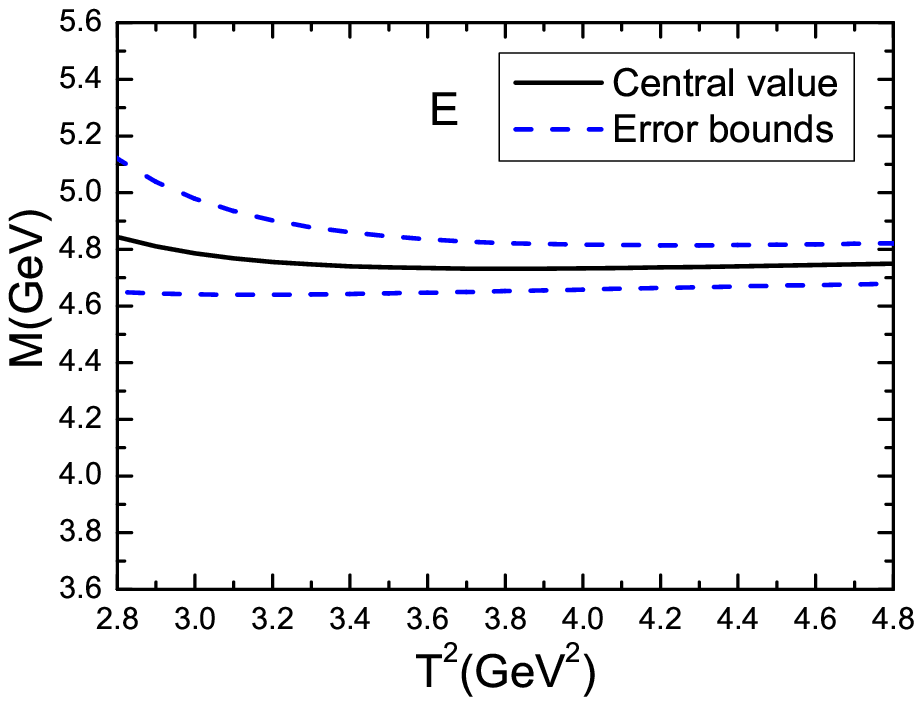}
 \includegraphics[totalheight=5cm,width=7cm]{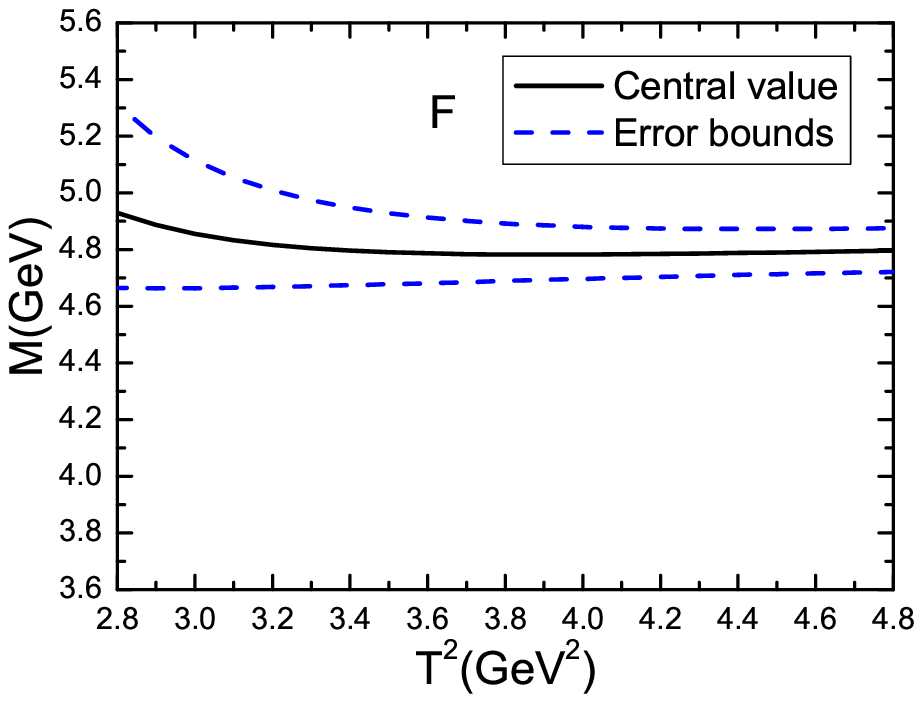}
         \caption{ The masses with variations of the Borel parameters, where  the $A$, $B$, $C$, $D$, $E$ and $F$ denote the tetraquark states  $cc\bar{u}\bar{d}\,(2^+)$, $cc\bar{u}\bar{s}\,(2^+)$, $cc\bar{s}\bar{s}\,(2^+)$, $cc\bar{u}\bar{d}\,(1^-)$, $cc\bar{u}\bar{s}\,(1^-)$ and $cc\bar{s}\bar{s}\,(1^-)$, respectively.   }
\end{figure}

\begin{figure}
 \centering
 \includegraphics[totalheight=5cm,width=7cm]{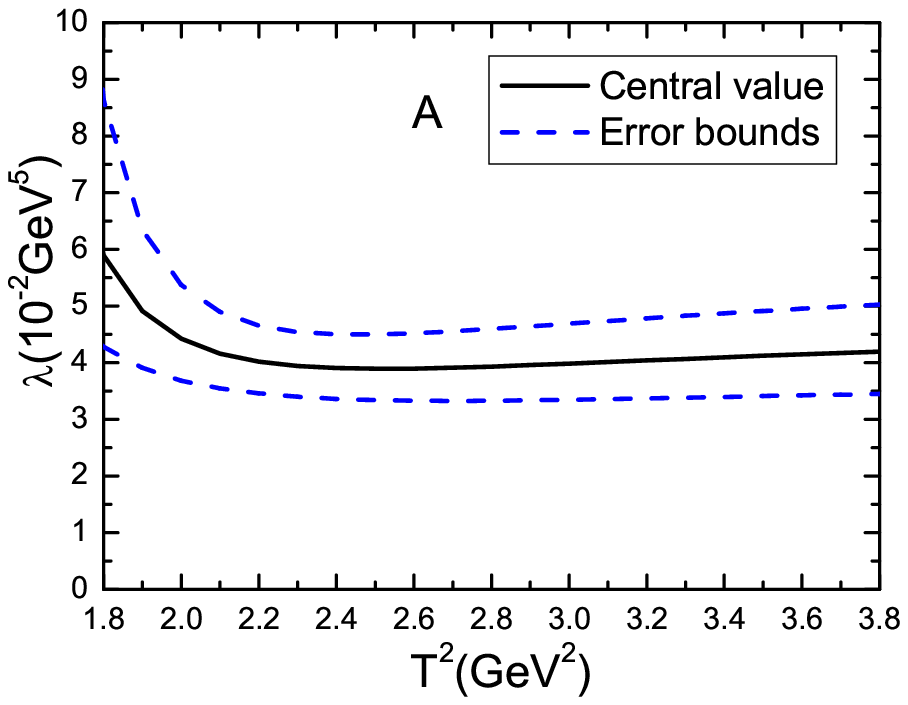}
 \includegraphics[totalheight=5cm,width=7cm]{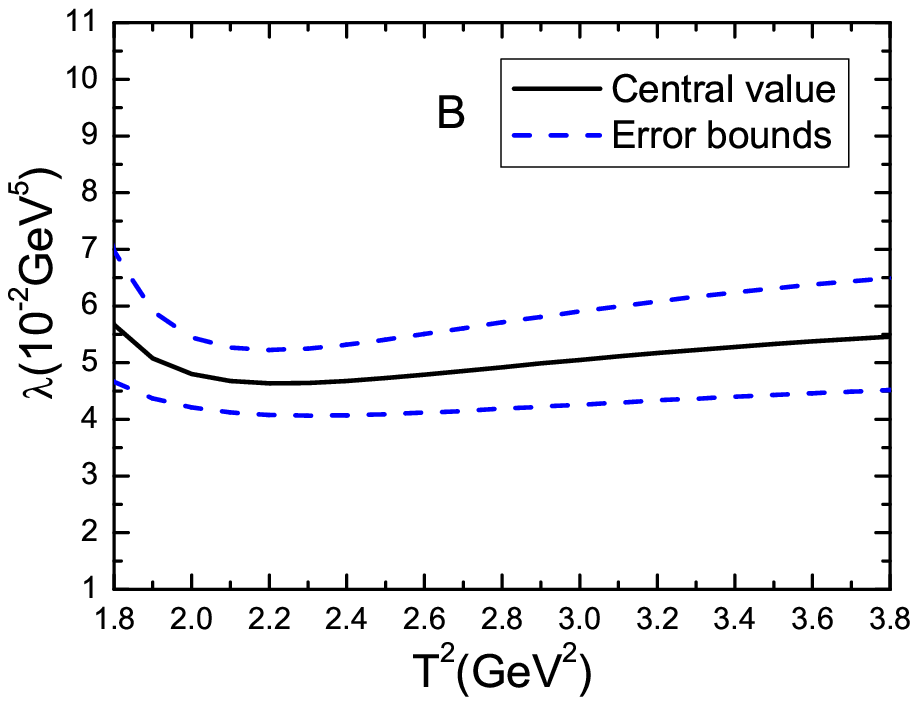}
 \includegraphics[totalheight=5cm,width=7cm]{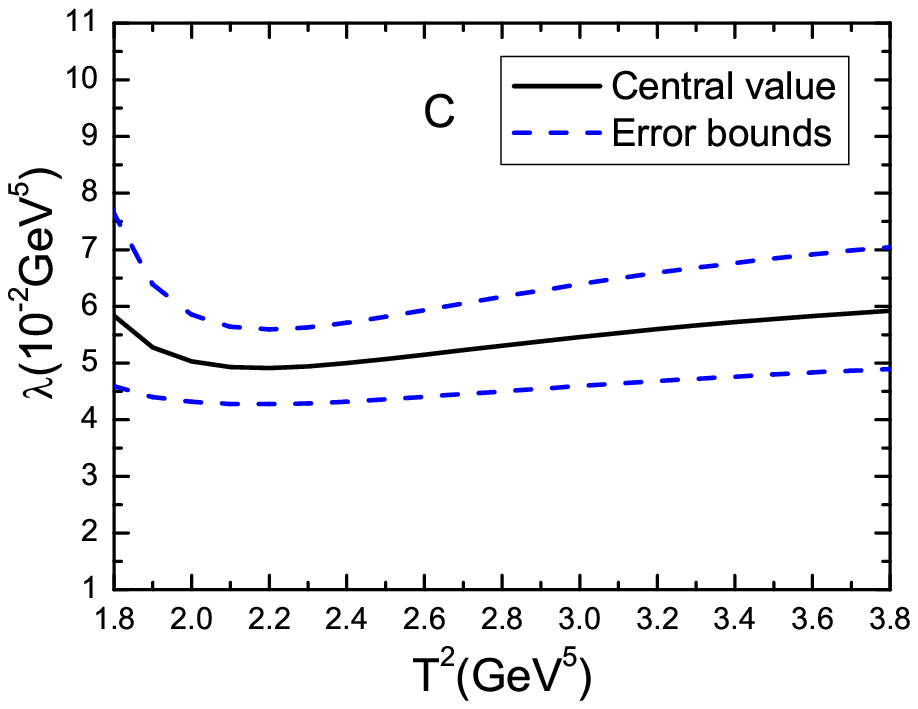}
 \includegraphics[totalheight=5cm,width=7cm]{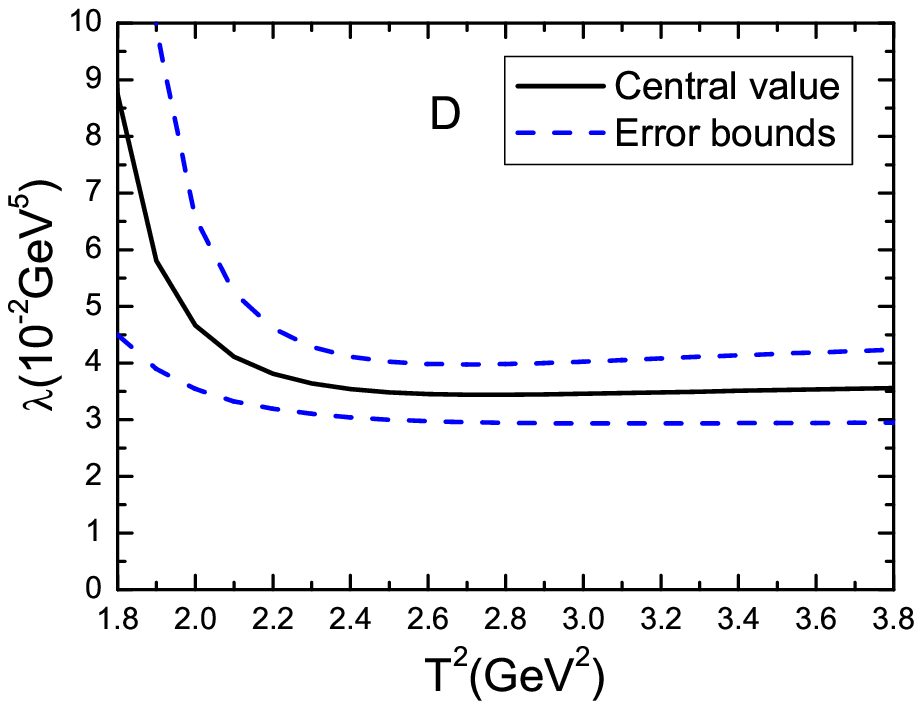}
 \includegraphics[totalheight=5cm,width=7cm]{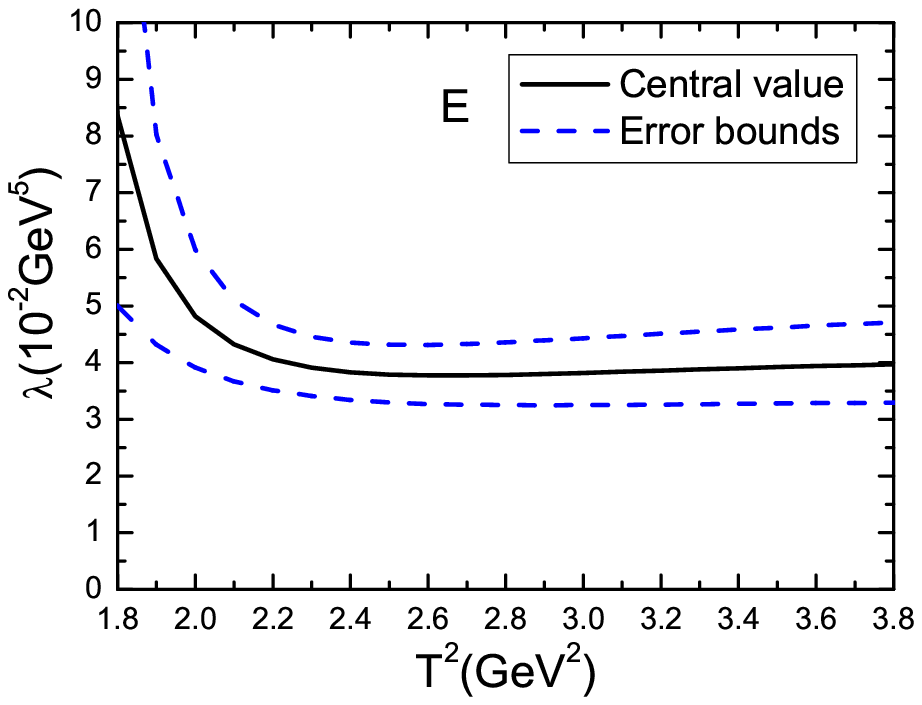}
 \includegraphics[totalheight=5cm,width=7cm]{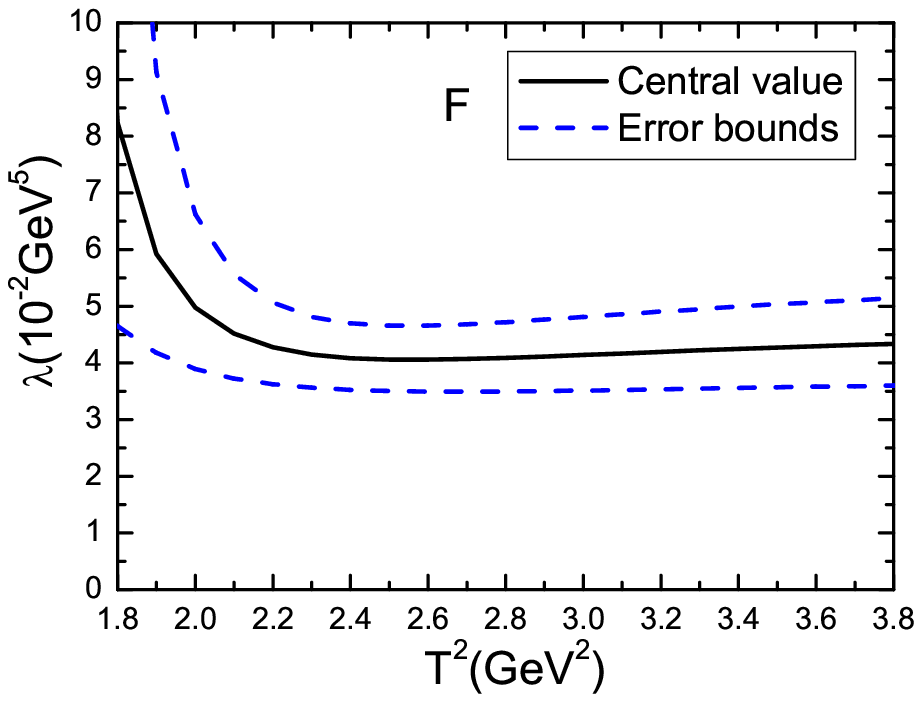}
         \caption{ The pole residues  with variations of the Borel parameters, where  the $A$, $B$, $C$, $D$, $E$ and $F$ denote the tetraquark states  $cc\bar{u}\bar{d}\,(0^+)$, $cc\bar{u}\bar{s}\,(0^+)$, $cc\bar{s}\bar{s}\,(0^+)$, $cc\bar{u}\bar{d}\,(1^+)$, $cc\bar{u}\bar{s}\,(1^+)$ and $cc\bar{s}\bar{s}\,(1^+)$, respectively.   }
\end{figure}

\begin{figure}
 \centering
 \includegraphics[totalheight=5cm,width=7cm]{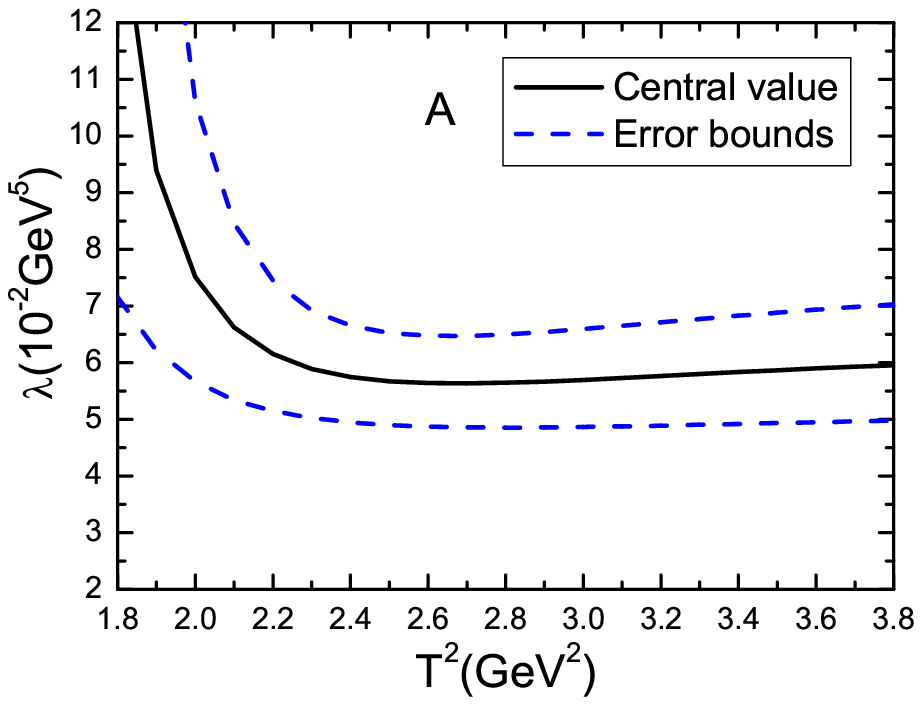}
 \includegraphics[totalheight=5cm,width=7cm]{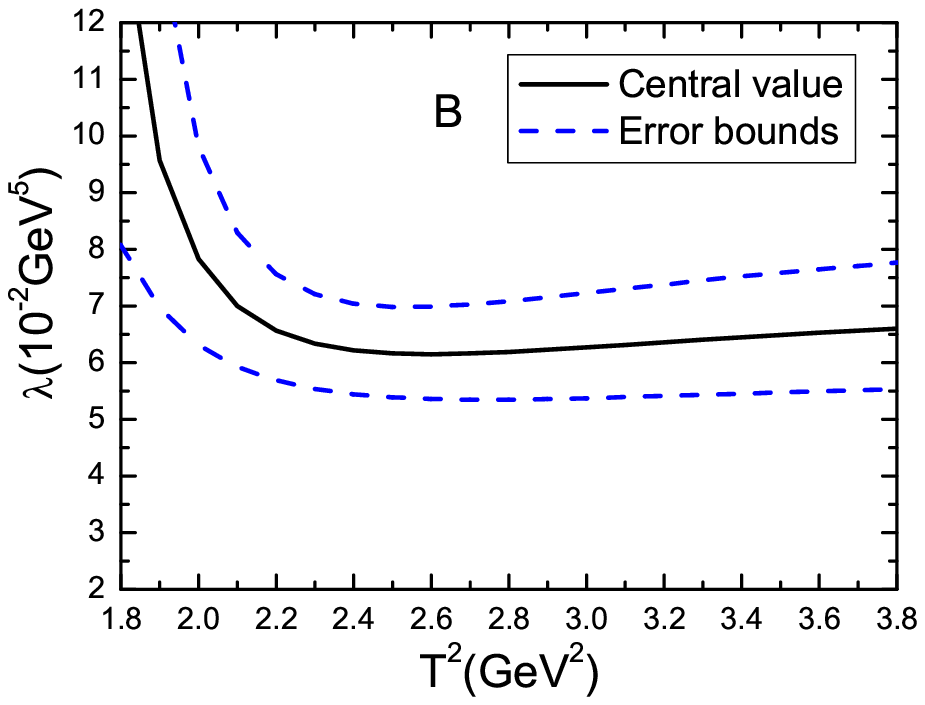}
 \includegraphics[totalheight=5cm,width=7cm]{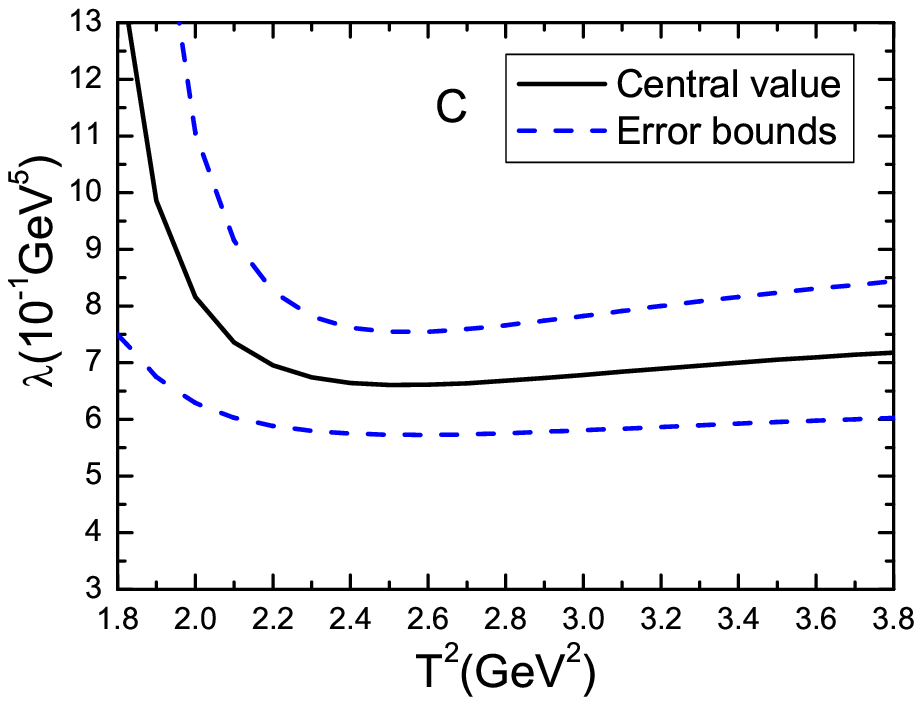}
 \includegraphics[totalheight=5cm,width=7cm]{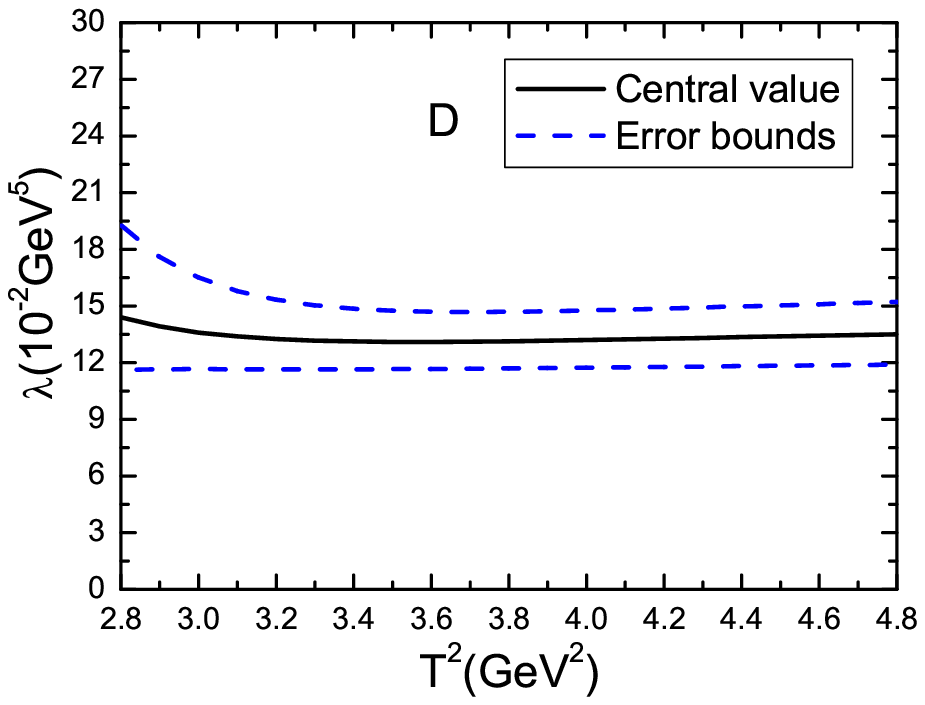}
 \includegraphics[totalheight=5cm,width=7cm]{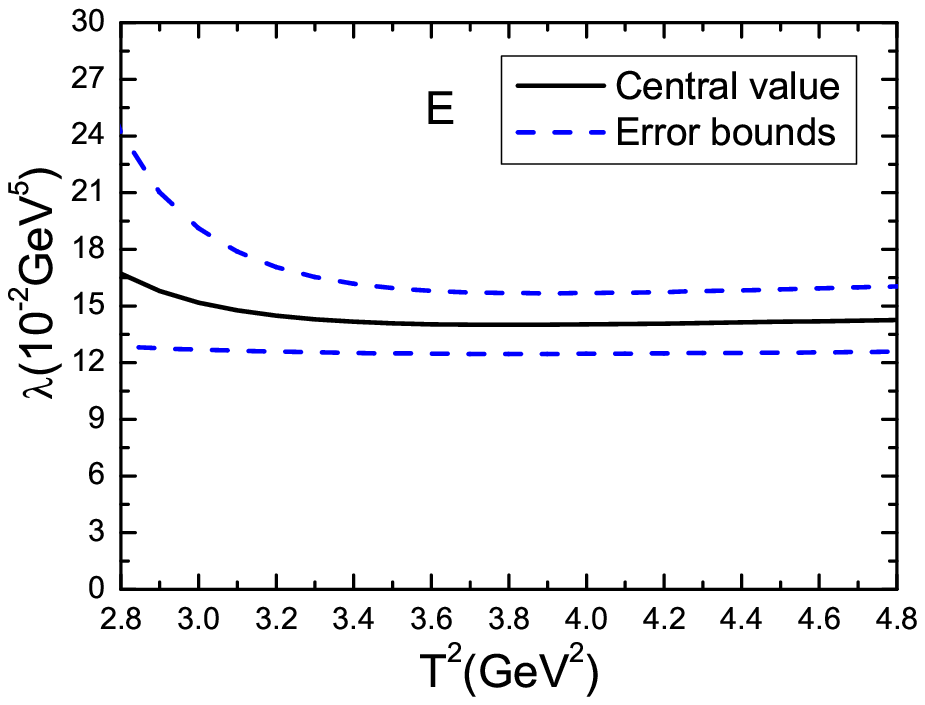}
 \includegraphics[totalheight=5cm,width=7cm]{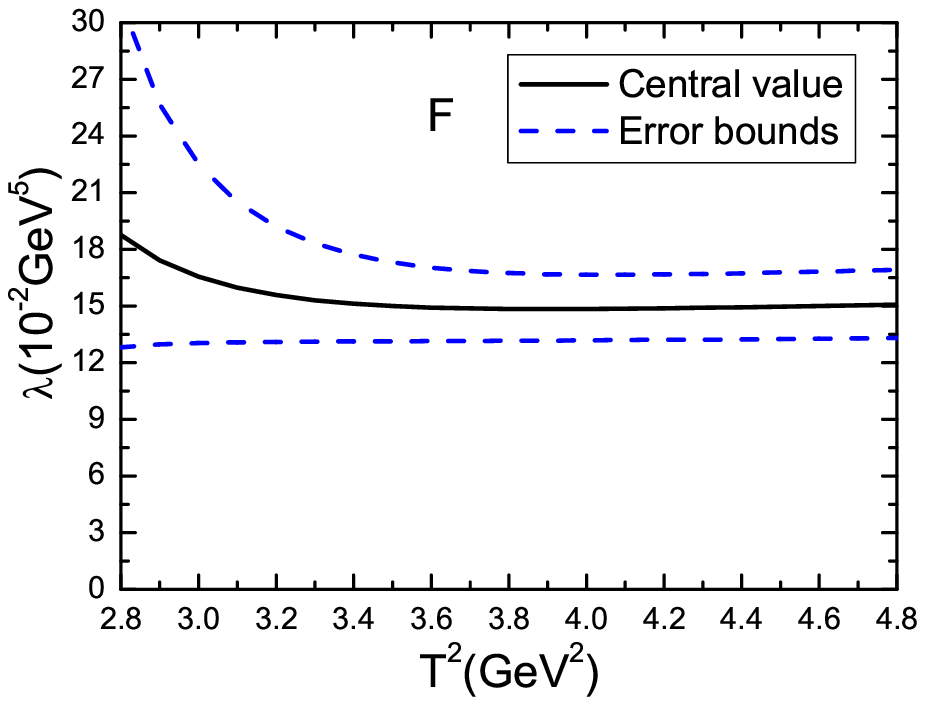}
         \caption{ The pole residues  with variations of the Borel parameters, where  the $A$, $B$, $C$, $D$, $E$ and $F$ denote the tetraquark states  $cc\bar{u}\bar{d}\,(2^+)$, $cc\bar{u}\bar{s}\,(2^+)$, $cc\bar{s}\bar{s}\,(2^+)$, $cc\bar{u}\bar{d}\,(1^-)$, $cc\bar{u}\bar{s}\,(1^-)$ and $cc\bar{s}\bar{s}\,(1^-)$, respectively.  }
\end{figure}

In Ref.\cite{Wang-4025-CTP}, we tentatively assign the $Z_c(4020/4025)$   to be the $C\gamma_\mu\otimes\gamma_\nu C$ type hidden-charm axialvector tetraquark state, and  choose the current,
\begin{eqnarray}
J_{\mu\nu;c\bar{c}}(x)&=&\varepsilon^{ijk}\varepsilon^{imn}\left\{u^T_j(x)C\gamma_\mu c_k(x) \bar{d}_m(x)\gamma_\nu C \bar{c}^T_n(x)-u^T_j(x)C\gamma_\nu c_k(x)\bar{d}_m(x)\gamma_\mu C \bar{c}^T_n(x) \right\} \, , \nonumber\\
\end{eqnarray}
to study it with the QCD sum rules.
In Ref.\cite{Wang-QQ-A}, we choose the axialvector current $J_{\mu;cc}(x)$,
\begin{eqnarray}
J_{\mu;cc}(x)&=&\varepsilon^{ijk}\varepsilon^{imn} \, c^{T}_j(x)C\gamma_\mu c_k(x) \,\bar{u}_m(x)\gamma_5C \bar{d}^T_n(x) \, ,
\end{eqnarray}
to study the $C\gamma_\mu\otimes\gamma_5C$ type   doubly charmed tetraquark state with the QCD sum rules.
In this article, we choose the axialvector current $J_{\mu\nu;\bar{u}\bar{d};1}(x)$,
\begin{eqnarray}
J_{\mu\nu;\bar{u}\bar{d};1}(x)&=&\varepsilon^{ijk}\varepsilon^{imn} \,\left[ c^{T}_j(x)C\gamma_\mu c_k(x) \,\bar{u}_m(x)\gamma_\nu C \bar{d}^T_n(x)-c^{T}_j(x)C\gamma_\nu c_k(x) \,\bar{u}_m(x)\gamma_\mu C \bar{d}^T_n(x) \right] \, , \nonumber\\
\end{eqnarray}
to study the $C\gamma_\mu\otimes\gamma_\nu C$ type doubly charmed tetraquark state.

In Fig.6, we plot
the masses  of the $C\gamma_\mu\otimes \gamma_\nu C$ type axialvector tetraquark state $c\bar{c}u\bar{d}$,  $C\gamma_\mu\otimes \gamma_5 C$ type axialvector tetraquark state $cc\bar{u}\bar{d}$ and $C\gamma_\mu\otimes \gamma_\nu C$ type axialvector tetraquark state $cc\bar{u}\bar{d}$ with variations of the Borel parameter $T^2$ for the energy scale $\mu=1.3\,\rm{GeV}$ and continuum threshold parameter $\sqrt{s_0}=4.45\,\rm{GeV}$. From the figure, we can see that the mass of the axialvector hidden-charm tetraquark state is $0.1\,\rm{GeV}$ larger than the ones of the corresponding  axialvector doubly charmed tetraquark states, while the $C\gamma_\mu\otimes \gamma_5 C$ type  and $C\gamma_\mu\otimes \gamma_\nu C$ type axialvector tetraquark states $cc\bar{u}\bar{d}$ have almost  degenerate  masses. In Ref.\cite{Wang-4025-CTP}, we observe that the calculations based on the QCD sum rules support that the $Z_c(4020/4025)$ can be assigned to be the axialvector hidden-charm tetraquark state. So the $C\gamma_\mu\otimes \gamma_5 C$ type  and $C\gamma_\mu\otimes \gamma_\nu C$ type axialvector tetraquark states $cc\bar{u}\bar{d}$ have the masses about $3.9\,\rm{GeV}$, the present predictions are reasonable. In Ref.\cite{Wang-QQ-A} and present work, we observe that we can choose a universal effective $c$-quark mass ${\mathbb{M}}_c=1.84\,\rm{GeV}$ to determine the energy scales of the QCD spectral densities in a consistent way, which leads to the energy scale $\mu=1.3\,\rm{GeV}$ for the QCD spectral density of the $C\gamma_\mu\otimes \gamma_5 C$ type tetraquark state $cc\bar{u}\bar{d}$. If we choose a slightly different energy scale $\mu=1.4\,\rm{GeV}$ (which corresponds  to a non-universal value ${\mathbb{M}}_c=1.82\,\rm{GeV}$) and  a slightly  different threshold parameter, we can obtain the lowest mass $3.85 \pm 0.09\, \rm{GeV}$, which is also shown in Table 1 in Ref.\cite{Wang-QQ-A}. In this article, we prefer the universal effective $c$-quark mass ${\mathbb{M}}_c=1.84\,\rm{GeV}$.

The centroids  of the masses of the $C\gamma_\mu\otimes \gamma_\nu C$ type tetraquark states are
\begin{eqnarray}
M_{C\gamma_\mu\otimes \gamma_\nu C}(cc\bar{u}\bar{d})&=&\frac{M_{cc\bar{u}\bar{d};0^+}+3M_{cc\bar{u}\bar{d};1^+}+5M_{cc\bar{u}\bar{d};2^+}}{9}=3.92\,\rm{GeV}\, , \nonumber\\
M_{C\gamma_\mu\otimes \gamma_\nu C}(cc\bar{u}\bar{s})&=&\frac{M_{cc\bar{u}\bar{s};0^+}+3M_{cc\bar{u}\bar{s};1^+}+5M_{cc\bar{u}\bar{s};2^+}}{9}=3.99\,\rm{GeV}\, , \nonumber\\
M_{C\gamma_\mu\otimes \gamma_\nu C}(cc\bar{s}\bar{s})&=&\frac{M_{cc\bar{s}\bar{s};0^+}+3M_{cc\bar{s}\bar{s};1^+}+5M_{cc\bar{s}\bar{s};2^+}}{9}=4.04\,\rm{GeV}\, ,
\end{eqnarray}
which are slightly larger than the centroids of the masses of the corresponding $C\gamma_\mu\otimes \gamma_5 C$ type tetraquark states,
\begin{eqnarray}
M_{C\gamma_\mu\otimes \gamma_5 C}(cc\bar{u}\bar{d})&=&3.90\,\rm{GeV}\, , \nonumber\\
M_{C\gamma_\mu\otimes \gamma_5 C}(cc\bar{u}\bar{s})&=&3.95\,\rm{GeV}\, ,
\end{eqnarray}
so the ground states are the $C\gamma_\mu\otimes \gamma_5 C$ type tetraquark states, which is consistent with our naive expectation that the axialvector (anti)diquarks have larger masses than the corresponding scalar (anti)diquarks. The lowest centroids $M_{cc\bar{u}\bar{d};0^+}=3.87\,\rm{GeV}$ and $M_{cc\bar{u}\bar{s};0^+}=3.94\,\rm{GeV}$ originate from the spin splitting,  in other words, the spin-spin interaction between the doubly heavy diquark and the light antidiquark. In fact, the predicted masses have uncertainties, the centroids  of the masses are not the super values, all values within uncertainties make sense.

In Ref.\cite{Eichten-Quigg},  Eichten and Quigg obtain the masses $M=4146\,\rm{MeV}$, $4167\,\rm{MeV}$ and $4210\,\rm{MeV}$ for the $C\gamma_\mu\otimes \gamma_\nu C$ type axialvector tetraquark states  $cc\bar{u}\bar{d}$, $cc\bar{u}\bar{s}$ and $cc\bar{s}\bar{s}$ respectively,   which are about $0.20-0.25\,\rm{GeV}$ larger than the central values of the present predictions. For the $C\gamma_\mu\otimes \gamma_5 C$ type   axialvector tetraquark state $cc\bar{u}\bar{d}$, Eichten and Quigg obtain the mass $M=3978\,\rm{MeV}$ \cite{Eichten-Quigg}, which is $0.1\,\rm{GeV}$ larger than the value $3882\,\rm{MeV}$ obtained by Karliner and  Rosner based on a simple potential quark model \cite{Karliner-Rosner}. The present predictions are consistent with the value $3882\,\rm{MeV}$ obtained by Karliner and  Rosner.

The doubly charmed tetraquark states   with the $J^P=0^+$, $1^+$ and $2^+$ lie near the corresponding  charmed meson pair thresholds, the decays to the  charmed meson   pairs are  Okubo-Zweig-Iizuka  super-allowed,
\begin{eqnarray}
Z_{cc\bar{u}\bar{d};0^+} &\to& D^0D^{+}\, , \nonumber\\
Z_{cc\bar{u}\bar{s};0^+} &\to& D^0D_s^{+}\, , \nonumber\\
Z_{cc\bar{s}\bar{s};0^+} &\to& D^+_s D_s^{+}\, , \nonumber\\
Z_{cc\bar{u}\bar{d};1^+} &\to& D^0D^{*+}\, , \,\, D^+D^{*0}\, , \nonumber\\
Z_{cc\bar{u}\bar{s};1^+} &\to& D^0D_s^{*+}\, , \,\, D^+_sD^{*0}\, , \nonumber\\
Z_{cc\bar{s}\bar{s};1^+} &\to& D^+_sD_s^{*+}\, , \nonumber\\
Z_{cc\bar{u}\bar{d};2^+} &\to& D^0D^{+}\, , \,\,D^{*0}D^{*+}\, , \nonumber\\
Z_{cc\bar{u}\bar{s};2^+} &\to& D^0D_s^{+}\, , \nonumber\\
Z_{cc\bar{s}\bar{s};2^+} &\to& D^+_s D_s^{+}\, ,
\end{eqnarray}
but the available  phase spaces are very small, the  decays are kinematically depressed, the  doubly charmed tetraquark states with the $J^P=0^+$, $1^+$ and $2^+$
maybe have small widths. On the other hand,  the doubly charmed tetraquark states   with the $J^P=1^-$ lie above  the corresponding  charmed meson pair thresholds, the decays to the  charmed meson   pairs are  Okubo-Zweig-Iizuka  super-allowed,
\begin{eqnarray}
Y_{cc\bar{u}\bar{d};1^-} &\to& D^0D^{+}\, , \,\,D^0D^{*+}\, , \,\, D^+D^{*0}\, , \nonumber\\
Y_{cc\bar{u}\bar{s};1^-} &\to& D^0D_s^{+}\, , \,\,D^0D_s^{*+}\, , \,\, D^+_sD^{*0}\, , \nonumber\\
Y_{cc\bar{s}\bar{s};1^-} &\to& D^+_sD_s^{+}\, , \,\, D^+_sD_s^{*+}\, ,
\end{eqnarray}
 the available  phase spaces are large, the  decays are kinematically facilitated, the  doubly charmed tetraquark states with the $J^P=1^-$ should
 have large widths.
 We can search for the doubly charmed tetraquark states in those decay channels in the future.

\begin{table}
\begin{center}
\begin{tabular}{|c|c|c|c|c|c|c|c|}\hline\hline
                           &$T^2(\rm{GeV}^2)$   &$\sqrt{s_0}(\rm{GeV})$   &$\mu(\rm{GeV})$  &pole          &$M(\rm{GeV})$  &$\lambda(\rm{GeV}^5)$ \\ \hline

$cc\bar{u}\bar{d}(0^+)$   &$2.4-2.8$           &$4.40\pm0.10$             &1.2              &$(38-63)\%$   &$3.87\pm0.09$  &$(3.90\pm0.63)\times10^{-2}$   \\ \hline

$cc\bar{u}\bar{s}(0^+)$   &$2.6-3.0$           &$4.50\pm0.10$             &1.3              &$(38-62)\%$   &$3.94\pm0.10$  &$(4.92\pm0.89)\times10^{-2}$   \\ \hline

$cc\bar{s}\bar{s}(0^+)$   &$2.6-3.0$           &$4.55\pm0.10$             &1.3              &$(39-63)\%$   &$3.99\pm0.10$  &$(5.31\pm0.99)\times10^{-2}$   \\ \hline

$cc\bar{u}\bar{d}(1^+)$   &$2.6-3.0$           &$4.45\pm0.10$             &1.3              &$(39-62)\%$   &$3.90\pm0.09$  &$(3.44\pm0.54)\times10^{-2}$   \\ \hline

$cc\bar{u}\bar{s}(1^+)$   &$2.6-3.0$           &$4.50\pm0.10$             &1.3              &$(40-64)\%$   &$3.96\pm0.08$  &$(3.78\pm0.59)\times10^{-2}$   \\ \hline

$cc\bar{s}\bar{s}(1^+)$   &$2.7-3.1$           &$4.55\pm0.10$             &1.3              &$(39-62)\%$   &$4.02\pm0.09$  &$(4.11\pm0.68)\times10^{-2}$   \\ \hline

$cc\bar{u}\bar{d}(2^+)$   &$2.7-3.1$           &$4.50\pm0.10$             &1.4              &$(39-62)\%$   &$3.95\pm0.09$  &$(5.67\pm0.90)\times10^{-2}$   \\ \hline

$cc\bar{u}\bar{s}(2^+)$   &$2.8-3.2$           &$4.55\pm0.10$             &1.4              &$(38-60)\%$   &$4.01\pm0.09$  &$(6.27\pm1.02)\times10^{-2}$   \\ \hline

$cc\bar{s}\bar{s}(2^+)$   &$2.8-3.2$           &$4.60\pm0.10$             &1.4              &$(39-61)\%$   &$4.06\pm0.09$  &$(6.78\pm1.12)\times10^{-2}$   \\ \hline

$cc\bar{u}\bar{d}(1^-)$   &$3.3-3.9$           &$5.20\pm0.10$             &2.9              &$(50-73)\%$   &$4.66\pm0.10$  &$(1.31\pm0.17)\times10^{-1}$   \\ \hline

$cc\bar{u}\bar{s}(1^-)$   &$3.4-4.0$           &$5.25\pm0.10$             &2.9              &$(49-71)\%$   &$4.73\pm0.11$  &$(1.40\pm0.19)\times10^{-1}$   \\ \hline

$cc\bar{s}\bar{s}(1^-)$   &$3.7-4.3$           &$5.30\pm0.10$             &2.9              &$(49-72)\%$   &$4.78\pm0.11$  &$(1.48\pm0.19)\times10^{-1}$   \\ \hline
 \hline
\end{tabular}
\end{center}
\caption{ The Borel parameters (Borel windows), continuum threshold parameters, optimal  energy scales, pole contributions,   masses and pole residues for the doubly charmed   tetraquark states. }
\end{table}

\begin{figure}
 \centering
 \includegraphics[totalheight=8cm,width=10cm]{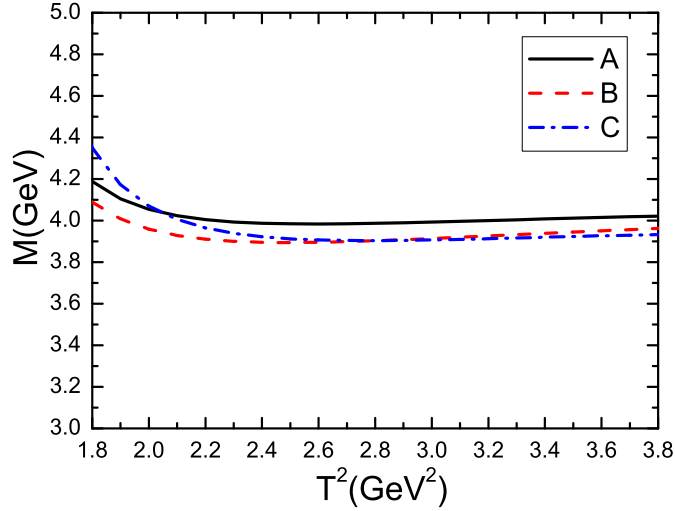}
         \caption{ The masses  of the axialvector tetraquark states with variations of the Borel parameter $T^2$ for the energy scale $\mu=1.3\,\rm{GeV}$ and continuum threshold parameter $\sqrt{s_0}=4.45\,\rm{GeV}$, where  the $A$, $B$  and  $C$ denote the  $C\gamma_\mu\otimes \gamma_\nu C$ type tetraquark state $c\bar{c}u\bar{d}$,  $C\gamma_\mu\otimes \gamma_5 C$ type tetraquark state $cc\bar{u}\bar{d}$ and $C\gamma_\mu\otimes \gamma_\nu C$ type tetraquark state $cc\bar{u}\bar{d}$, respectively.   }
\end{figure}

\section{Conclusion}
In this article, we construct the axialvector-diquark-axialvector-antidiquark type currents to interpolate the  scalar, axialvector, vector, tensor doubly charmed  tetraquark states, and study them with the QCD sum rules in a systematic way. In calculations, we carry out the operator product expansion up to the vacuum condensates of dimension 10 consistently, then obtain the QCD spectral densities through dispersion relation, and extract the masses and pole residues in the Borel windows at the optimal energy scales of the QCD spectral densities, which are determined by  the energy scale formula with the refitted  effective charm quark mass ${\mathbb{M}}_c$.
In the Borel windows,  the pole dominance is satisfied  and  the operator product expansion is  well convergent, so we expect to make reliable  predictions.   We can search for those doubly charmed  tetraquark states in the  Okubo-Zweig-Iizuka  super-allowed strong decays to the charmed-meson pairs in the future.

\section*{Appendix}

The explicit expressions of the QCD spectral densities $\rho_{\bar{u}\bar{d};0}(s)$, $\rho_{\bar{u}\bar{s};0}(s)$, $\rho_{\bar{s}\bar{s};0}(s)$, $\rho_{\bar{u}\bar{d};1;A}(s)$, $\rho_{\bar{u}\bar{s};1;A}(s)$, $\rho_{\bar{s}\bar{s};1;A}(s)$, $\rho_{\bar{u}\bar{d};1;V}(s)$, $\rho_{\bar{u}\bar{s};1;V}(s)$, $\rho_{\bar{s}\bar{s};1;V}(s)$,  $\rho_{\bar{u}\bar{d};2}(s)$, $\rho_{\bar{u}\bar{s};2}(s)$, $\rho_{\bar{s}\bar{s};2}(s)$,

\begin{eqnarray}
\rho_{\bar{u}\bar{s};0}(s)&=&\rho_{0;0}(s)+\rho_{3;0}(s)+\rho_{4;0}(s)+\rho_{5;0}(s)+\rho_{6;0}(s)+\rho_{8;0}(s)+\rho_{10;0}(s)\, , \nonumber\\
\rho_{\bar{u}\bar{s};1;A}(s)&=&\rho_{0;1;A}(s)+\rho_{3;1;A}(s)+\rho_{4;1;A}(s)+\rho_{5;1;A}(s)+\rho_{6;1;A}(s)+\rho_{8;1;A}(s)+\rho_{10;1;A}(s)\, , \nonumber\\
\rho_{\bar{u}\bar{s};1;V}(s)&=&\rho_{0;1;V}(s)+\rho_{3;1;V}(s)+\rho_{4;1;V}(s)+\rho_{5;1;V}(s)+\rho_{6;1;V}(s)+\rho_{8;1;V}(s)+\rho_{10;1;V}(s)\, , \nonumber\\
\rho_{\bar{u}\bar{s};2}(s)&=&\rho_{0;2}(s)+\rho_{3;2}(s)+\rho_{4;2}(s)+\rho_{5;2}(s)+\rho_{6;2}(s)+\rho_{8;2}(s)+\rho_{10;2}(s)\, ,
\end{eqnarray}

\begin{eqnarray}
\rho_{\bar{u}\bar{d};0}(s)&=&\rho_{\bar{u}\bar{s};0}(s)\mid_{m_s \to 0, \, \langle\bar{s}s\rangle \to \langle\bar{q}q\rangle,\,\langle\bar{s}g_s\sigma Gs\rangle \to \langle\bar{q}g_s\sigma Gq\rangle }\, , \nonumber\\
\rho_{\bar{u}\bar{d};1;A}(s)&=&\rho_{\bar{u}\bar{s};1;A}(s)\mid_{m_s \to 0, \, \langle\bar{s}s\rangle \to \langle\bar{q}q\rangle,\,\langle\bar{s}g_s\sigma Gs\rangle \to \langle\bar{q}g_s\sigma Gq\rangle }\, , \nonumber\\
\rho_{\bar{u}\bar{d};1;V}(s)&=&\rho_{\bar{u}\bar{s};1;V}(s)\mid_{m_s \to 0, \, \langle\bar{s}s\rangle \to \langle\bar{q}q\rangle,\,\langle\bar{s}g_s\sigma Gs\rangle \to \langle\bar{q}g_s\sigma Gq\rangle }\, , \nonumber\\
\rho_{\bar{u}\bar{d};2}(s)&=&\rho_{\bar{u}\bar{s};2}(s)\mid_{m_s \to 0, \, \langle\bar{s}s\rangle \to \langle\bar{q}q\rangle,\,\langle\bar{s}g_s\sigma Gs\rangle \to \langle\bar{q}g_s\sigma Gq\rangle }\, ,
\end{eqnarray}

\begin{eqnarray}
\rho_{\bar{s}\bar{s};0}(s)&=&\rho_{\bar{u}\bar{s};0}(s)\mid_{m_s \to 2m_s, \, \langle\bar{q}q\rangle \to \langle\bar{s}s\rangle,\,\langle\bar{q}g_s\sigma Gq\rangle \to \langle\bar{s}g_s\sigma Gs\rangle }\, , \nonumber\\
\rho_{\bar{s}\bar{s};1;A}(s)&=&\rho_{\bar{u}\bar{s};1;A}(s)\mid_{m_s \to 2m_s, \, \langle\bar{q}q\rangle \to \langle\bar{s}s\rangle,\,\langle\bar{q}g_s\sigma Gq\rangle \to \langle\bar{s}g_s\sigma Gs\rangle }\, , \nonumber\\
\rho_{\bar{s}\bar{s};1;V}(s)&=&\rho_{\bar{u}\bar{s};1;V}(s)\mid_{m_s \to 2m_s, \, \langle\bar{q}q\rangle \to \langle\bar{s}s\rangle,\,\langle\bar{q}g_s\sigma Gq\rangle \to \langle\bar{s}g_s\sigma Gs\rangle }\, , \nonumber\\
\rho_{\bar{s}\bar{s};2}(s)&=&\rho_{\bar{u}\bar{s};2}(s)\mid_{m_s \to 2m_s, \, \langle\bar{q}q\rangle \to \langle\bar{s}s\rangle,\,\langle\bar{q}g_s\sigma Gq\rangle \to \langle\bar{s}g_s\sigma Gs\rangle }\, ,
\end{eqnarray}

\begin{eqnarray}
\rho_{0;0}(s)&=&\frac{1}{64\pi^{6}}\int_{y_i}^{y_f}dy \int_{z_i}^{1-y}dz\,yz(1-y-z)^2\left(s-\overline{m}_c^2\right)^3\left(3s-\overline{m}_c^2\right) \nonumber\\
&&+\frac{m_c^2}{64\pi^{6}}\int_{y_i}^{y_f}dy \int_{z_i}^{1-y}dz\,(1-y-z)^2\left(s-\overline{m}_c^2\right)^3 \, ,
\end{eqnarray}

\begin{eqnarray}
\rho_{3;0}(s)&=&-\frac{m_s\left[\langle\bar{q}q\rangle-\langle\bar{s}s\rangle \right]}{4\pi^{4}}\int_{y_i}^{y_f}dy \int_{z_i}^{1-y}dz\,yz\left(s-\overline{m}_c^2\right)\left(2s-\overline{m}_c^2\right)  \nonumber\\
&&-\frac{m_s m_c^2\left[4\langle\bar{q}q\rangle-\langle\bar{s}s\rangle \right]}{8\pi^{4}}\int_{y_i}^{y_f}dy \int_{z_i}^{1-y}dz\, \left(s-\overline{m}_c^2\right) \, ,
\end{eqnarray}

\begin{eqnarray}
\rho_{4;0}(s)&=&-\frac{m_c^2}{96\pi^{4}}\langle \frac{\alpha_sGG}{\pi}\rangle\int_{y_i}^{y_f}dy \int_{z_i}^{1-y}dz\,\left(\frac{z}{y^2}+\frac{y}{z^2} \right)(1-y-z)^2\left(3s-2\overline{m}_c^2\right)  \nonumber\\
&&-\frac{m_c^4}{192\pi^{4}}\langle \frac{\alpha_sGG}{\pi}\rangle\int_{y_i}^{y_f}dy \int_{z_i}^{1-y}dz\,\left(\frac{1}{y^3}+\frac{1}{z^3} \right)(1-y-z)^2  \nonumber\\
&&+\frac{m_c^2}{64\pi^{4}}\langle \frac{\alpha_sGG}{\pi}\rangle\int_{y_i}^{y_f}dy \int_{z_i}^{1-y}dz\,\left[\left(\frac{1}{y^2}+\frac{1}{z^2} \right)(1-y-z)^2-1 \right] \left(s-\overline{m}_c^2\right)  \, ,
\end{eqnarray}

\begin{eqnarray}
\rho_{5;0}(s)&=& \frac{m_s\left[3\langle\bar{q}g_s\sigma Gq\rangle-2\langle\bar{s}g_s\sigma Gs\rangle \right]}{48\pi^{4}}\int_{y_i}^{y_f}dy  \, y(1-y) \left(3s-2\widetilde{m}_c^2\right)   \nonumber\\
&&+ \frac{m_s m_c^2\left[6\langle\bar{q}g_s\sigma Gq\rangle-\langle\bar{s}g_s\sigma Gs\rangle \right]}{48\pi^{4}}\int_{y_i}^{y_f}dy      \nonumber\\
&&- \frac{m_s\langle\bar{q}g_s\sigma Gq\rangle}{64\pi^{4}}\int_{y_i}^{y_f}dy  \int_{z_i}^{1-y}dz\,(y+z) \left(3s-2\overline{m}_c^2\right)   \nonumber\\
&&- \frac{m_s m_c^2\langle\bar{q}g_s\sigma Gq\rangle}{32\pi^{4}}\int_{y_i}^{y_f}dy  \int_{z_i}^{1-y}dz\,\left(\frac{1}{y}+\frac{1}{z}\right)    \, ,
\end{eqnarray}

\begin{eqnarray}
\rho_{6;0}(s)&=&\frac{\langle\bar{q}q\rangle\langle\bar{s}s\rangle }{\pi^{2}}\int_{y_i}^{y_f}dy \,y(1-y)s\, ,
\end{eqnarray}

\begin{eqnarray}
\rho_{8;0}(s)&=&-\frac{\langle\bar{s}s\rangle\langle\bar{q}g_s\sigma Gq\rangle+\langle\bar{q}q\rangle\langle\bar{s}g_s\sigma Gs\rangle }{4\pi^{2}}\int_{y_i}^{y_f}dy \,y(1-y)\left[2+\left(2s+\frac{s^2}{T^2}\right)\,\delta\left(s-\widetilde{m}_c^2\right)\right]      \nonumber\\
&&+\frac{\langle\bar{s}s\rangle\langle\bar{q}g_s\sigma Gq\rangle+\langle\bar{q}q\rangle\langle\bar{s}g_s\sigma Gs\rangle }{48\pi^{2}}\int_{y_i}^{y_f}dy \, \left[2+3s\,\delta\left(s-\widetilde{m}_c^2\right)\right]      \, ,
\end{eqnarray}

\begin{eqnarray}
\rho_{10;0}(s)&=& \frac{ \langle\bar{q}g_s\sigma Gq\rangle  \langle\bar{s}g_s\sigma Gs\rangle }{16\pi^{2}}\int_{y_i}^{y_f}dy \, y(1-y)\,\left(2+\frac{2s}{T^2}+\frac{s^2}{T^4}+\frac{s^3}{T^6} \right) \delta\left(s-\widetilde{m}_c^2\right)  \nonumber\\
&& -\frac{ \langle\bar{q}g_s\sigma Gq\rangle  \langle\bar{s}g_s\sigma Gs\rangle }{96\pi^{2}}\int_{y_i}^{y_f}dy \, \,\left(2+\frac{2s}{T^2}+\frac{3s^2}{T^4} \right) \delta\left(s-\widetilde{m}_c^2\right)  \nonumber\\
&& +\frac{11 \langle\bar{q}g_s\sigma Gq\rangle  \langle\bar{s}g_s\sigma Gs\rangle }{768\pi^{2}}\int_{y_i}^{y_f}dy \, \,\left(1+\frac{5s}{T^2} \right) \delta\left(s-\widetilde{m}_c^2\right)  \, ,
\end{eqnarray}

\begin{eqnarray}
\rho_{0;1;A}(s)&=&\frac{1}{384\pi^{6}}\int_{y_i}^{y_f}dy \int_{z_i}^{1-y}dz\,yz(1-y-z)^3\left(s-\overline{m}_c^2\right)^2\left(21s^2-14s\overline{m}_c^2+\overline{m}_c^4\right) \nonumber\\
&&-\frac{1}{384\pi^{6}}\int_{y_i}^{y_f}dy \int_{z_i}^{1-y}dz\,yz(1-y-z)^2\left(s-\overline{m}_c^2\right)^3\left( 3s -\overline{m}_c^2\right) \nonumber\\
&&+\frac{m_c^2}{576\pi^{6}}\int_{y_i}^{y_f}dy \int_{z_i}^{1-y}dz\,(1-y-z)^3\left(s-\overline{m}_c^2\right)^2 \left( 7s -\overline{m}_c^2\right) \nonumber\\
&&+\frac{m_c^2}{192\pi^{6}}\int_{y_i}^{y_f}dy \int_{z_i}^{1-y}dz\,(1-y-z)^2\left(s-\overline{m}_c^2\right)^3 \, ,
\end{eqnarray}

\begin{eqnarray}
\rho_{3;1;A}(s)&=&\frac{m_s\langle\bar{s}s\rangle}{24\pi^{4}}\int_{y_i}^{y_f}dy \int_{z_i}^{1-y}dz\,yz (1-y-z)\left(25s^2-24s\overline{m}_c^2+3\overline{m}_c^4\right)  \nonumber\\
&&-\frac{m_s\langle\bar{q}q\rangle}{24\pi^{4}}\int_{y_i}^{y_f}dy \int_{z_i}^{1-y}dz\,yz \left(s-\overline{m}_c^2\right)\left( 5s -\overline{m}_c^2\right)  \nonumber\\
&&-\frac{m_s\left[2\langle\bar{q}q\rangle+\langle\bar{s}s\rangle\right]}{24\pi^{4}}\int_{y_i}^{y_f}dy \int_{z_i}^{1-y}dz\,yz \left(s-\overline{m}_c^2\right)\left( 2s -\overline{m}_c^2\right)  \nonumber\\
&&-\frac{m_s m_c^2\left[6\langle\bar{q}q\rangle-\langle\bar{s}s\rangle \right]}{24\pi^{4}}\int_{y_i}^{y_f}dy \int_{z_i}^{1-y}dz\, \left(s-\overline{m}_c^2\right) \nonumber\\
&&+\frac{m_s m_c^2\langle\bar{s}s\rangle }{24\pi^{4}}\int_{y_i}^{y_f}dy \int_{z_i}^{1-y}dz\,(1-y-z) \left(3s-\overline{m}_c^2\right) \, ,
\end{eqnarray}

\begin{eqnarray}
\rho_{4;1;A}(s)&=&-\frac{m_c^2}{288\pi^{4}}\langle \frac{\alpha_sGG}{\pi}\rangle\int_{y_i}^{y_f}dy \int_{z_i}^{1-y}dz\,\left(\frac{z}{y^2}+\frac{y}{z^2} \right)(1-y-z)^3\left[4s-\overline{m}_c^2+\frac{2s^2}{3}\delta \left(s-\overline{m}_c^2\right) \right]  \nonumber\\
&&+\frac{m_c^2}{576\pi^{4}}\langle \frac{\alpha_sGG}{\pi}\rangle\int_{y_i}^{y_f}dy \int_{z_i}^{1-y}dz\,\left(\frac{z}{y^2}+\frac{y}{z^2} \right)(1-y-z)^2\left(3s-2\overline{m}_c^2 \right)   \nonumber\\
&&-\frac{m_c^4}{864\pi^{4}}\langle \frac{\alpha_sGG}{\pi}\rangle\int_{y_i}^{y_f}dy \int_{z_i}^{1-y}dz\,\left(\frac{1}{y^3}+\frac{1}{z^3} \right)(1-y-z)^3 \left[\frac{1}{2}+s\,\delta \left(s-\overline{m}_c^2\right) \right] \nonumber\\
&&+\frac{m_c^2}{576\pi^{4}}\langle \frac{\alpha_sGG}{\pi}\rangle\int_{y_i}^{y_f}dy \int_{z_i}^{1-y}dz\,(1-y-z)\left[1+\left(\frac{1}{y^2}+\frac{1}{z^2} \right)(1-y-z)^2\right] \left(3s-\overline{m}_c^2\right)   \nonumber\\
&&-\frac{m_c^4}{576\pi^{4}}\langle \frac{\alpha_sGG}{\pi}\rangle\int_{y_i}^{y_f}dy \int_{z_i}^{1-y}dz\,\left(\frac{1}{y^3}+\frac{1}{z^3} \right)(1-y-z)^2 \nonumber\\
&&+\frac{m_c^2}{576\pi^{4}}\langle \frac{\alpha_sGG}{\pi}\rangle\int_{y_i}^{y_f}dy \int_{z_i}^{1-y}dz\,\left[\left(\frac{3}{y^2}+\frac{3}{z^2} \right)(1-y-z)^2 -5\right] \left(s-\overline{m}_c^2\right)   \nonumber\\
&&-\frac{1}{384\pi^{4}}\langle \frac{\alpha_sGG}{\pi}\rangle\int_{y_i}^{y_f}dy \int_{z_i}^{1-y}dz\, (1-y-z)^2 \left(s-\overline{m}_c^2\right)\left(2s-\overline{m}_c^2\right)    \nonumber\\
&&+\frac{1}{576\pi^{4}}\langle \frac{\alpha_sGG}{\pi}\rangle\int_{y_i}^{y_f}dy \int_{z_i}^{1-y}dz\, yz(1-y-z) \left(25s^2-24s\overline{m}_c^2+3\overline{m}_c^4\right)    \nonumber\\
&&-\frac{1}{576\pi^{4}}\langle \frac{\alpha_sGG}{\pi}\rangle\int_{y_i}^{y_f}dy \int_{z_i}^{1-y}dz\, yz \left(s-\overline{m}_c^2\right)\left(11s-4\overline{m}_c^2\right) \, ,
\end{eqnarray}

\begin{eqnarray}
\rho_{5;1;A}(s)&=&- \frac{m_s\langle\bar{s}g_s\sigma Gs\rangle}{24\pi^{4}}\int_{y_i}^{y_f}dy  \int_{z_i}^{1-y}dz\,yz \left[4s-\overline{m}_c^2+\frac{2s^2}{3}\delta \left(s-\overline{m}_c^2\right) \right]  \nonumber\\
&& +\frac{m_s\langle\bar{q}g_s\sigma Gq\rangle}{48\pi^{4}}\int_{y_i}^{y_f}dy  \, y(1-y) \left(3s-\widetilde{m}_c^2\right)   \nonumber\\
&& +\frac{m_s\left[3\langle\bar{q}g_s\sigma Gq\rangle+\langle\bar{s}g_s\sigma Gs\rangle\right]}{144\pi^{4}}\int_{y_i}^{y_f}dy  \, y(1-y) \left(3s-2\widetilde{m}_c^2\right)   \nonumber\\
&&- \frac{m_s m_c^2\langle\bar{s}g_s\sigma Gs\rangle}{72\pi^{4}}\int_{y_i}^{y_f}dy \int_{z_i}^{1-y}dz \, \left[\frac{1}{2}+s\,\delta \left(s-\overline{m}_c^2\right) \right]   \nonumber\\
&&+ \frac{m_s m_c^2\left[9\langle\bar{q}g_s\sigma Gq\rangle-\langle\bar{s}g_s\sigma Gs\rangle \right]}{144\pi^{4}}\int_{y_i}^{y_f}dy      \nonumber\\
&&- \frac{m_s\langle\bar{q}g_s\sigma Gq\rangle}{192\pi^{4}}\int_{y_i}^{y_f}dy  \int_{z_i}^{1-y}dz\,(y+z) \left(2s-\overline{m}_c^2\right)   \nonumber\\
&&- \frac{m_s m_c^2\langle\bar{q}g_s\sigma Gq\rangle}{192\pi^{4}}\int_{y_i}^{y_f}dy  \int_{z_i}^{1-y}dz\,\left(\frac{1}{y}+\frac{1}{z}\right)    \, ,
\end{eqnarray}

\begin{eqnarray}
\rho_{6;1;A}(s)&=&\frac{2\langle\bar{q}q\rangle\langle\bar{s}s\rangle }{3\pi^{2}}\int_{y_i}^{y_f}dy \,y(1-y)s \, ,
\end{eqnarray}

\begin{eqnarray}
\rho_{8;1;A}(s)&=&-\frac{\langle\bar{s}s\rangle\langle\bar{q}g_s\sigma Gq\rangle+\langle\bar{q}q\rangle\langle\bar{s}g_s\sigma Gs\rangle }{12\pi^{2}}\int_{y_i}^{y_f}dy \,y(1-y)\left[3+\left(4s+\frac{2s^2}{T^2}\right)\,\delta\left(s-\widetilde{m}_c^2\right)\right]      \nonumber\\
&&+\frac{\langle\bar{s}s\rangle\langle\bar{q}g_s\sigma Gq\rangle+\langle\bar{q}q\rangle\langle\bar{s}g_s\sigma Gs\rangle }{144\pi^{2}}\int_{y_i}^{y_f}dy \, \left[1+2s\,\delta\left(s-\widetilde{m}_c^2\right)\right]  \, ,
\end{eqnarray}

\begin{eqnarray}
\rho_{10;1;A}(s)&=& \frac{ \langle\bar{q}g_s\sigma Gq\rangle  \langle\bar{s}g_s\sigma Gs\rangle }{24\pi^{2}}\int_{y_i}^{y_f}dy \, y(1-y)\,\left(\frac{s}{T^2}+\frac{s^2}{T^4}+\frac{s^3}{T^6} \right) \delta\left(s-\widetilde{m}_c^2\right)  \nonumber\\
&& -\frac{ \langle\bar{q}g_s\sigma Gq\rangle  \langle\bar{s}g_s\sigma Gs\rangle }{288\pi^{2}}\int_{y_i}^{y_f}dy \, \,\left(\frac{s}{T^2}+\frac{2s^2}{T^4} \right) \delta\left(s-\widetilde{m}_c^2\right)  \nonumber\\
&& -\frac{11 \langle\bar{q}g_s\sigma Gq\rangle  \langle\bar{s}g_s\sigma Gs\rangle }{768\pi^{2}}\int_{y_i}^{y_f}dy \, \,\frac{s}{T^2} \delta\left(s-\widetilde{m}_c^2\right)  \, ,
\end{eqnarray}

\begin{eqnarray}
\rho_{0;1;V}(s)&=&\frac{1}{384\pi^{6}}\int_{y_i}^{y_f}dy \int_{z_i}^{1-y}dz\,yz(1-y-z)^3\left(s-\overline{m}_c^2\right)^2\left(21s^2-14s\overline{m}_c^2+\overline{m}_c^4\right) \nonumber\\
&&-\frac{1}{384\pi^{6}}\int_{y_i}^{y_f}dy \int_{z_i}^{1-y}dz\,yz(1-y-z)^2\left(s-\overline{m}_c^2\right)^3\left( 3s -\overline{m}_c^2\right) \nonumber\\
&&+\frac{m_c^2}{576\pi^{6}}\int_{y_i}^{y_f}dy \int_{z_i}^{1-y}dz\,(1-y-z)^3\left(s-\overline{m}_c^2\right)^2 \left( 7s -\overline{m}_c^2\right) \nonumber\\
&&-\frac{m_c^2}{96\pi^{6}}\int_{y_i}^{y_f}dy \int_{z_i}^{1-y}dz\,(1-y-z)^2\left(s-\overline{m}_c^2\right)^3 \, ,
\end{eqnarray}

\begin{eqnarray}
\rho_{3;1;V}(s)&=&\frac{m_s\langle\bar{s}s\rangle}{24\pi^{4}}\int_{y_i}^{y_f}dy \int_{z_i}^{1-y}dz\,yz (1-y-z)\left(25s^2-24s\overline{m}_c^2+3\overline{m}_c^4\right)  \nonumber\\
&&-\frac{m_s\langle\bar{q}q\rangle}{24\pi^{4}}\int_{y_i}^{y_f}dy \int_{z_i}^{1-y}dz\,yz \left(s-\overline{m}_c^2\right)\left( 5s -\overline{m}_c^2\right)  \nonumber\\
&&+\frac{m_s\left[4\langle\bar{q}q\rangle-\langle\bar{s}s\rangle\right]}{24\pi^{4}}\int_{y_i}^{y_f}dy \int_{z_i}^{1-y}dz\,yz \left(s-\overline{m}_c^2\right)\left( 2s -\overline{m}_c^2\right)  \nonumber\\
&&+\frac{m_s m_c^2\left[4\langle\bar{q}q\rangle-\langle\bar{s}s\rangle \right]}{12\pi^{4}}\int_{y_i}^{y_f}dy \int_{z_i}^{1-y}dz\, \left(s-\overline{m}_c^2\right) \nonumber\\
&&+\frac{m_s m_c^2\langle\bar{s}s\rangle }{24\pi^{4}}\int_{y_i}^{y_f}dy \int_{z_i}^{1-y}dz\,(1-y-z) \left(3s-\overline{m}_c^2\right) \, ,
\end{eqnarray}

\begin{eqnarray}
\rho_{4;1;V}(s)&=&-\frac{m_c^2}{288\pi^{4}}\langle \frac{\alpha_sGG}{\pi}\rangle\int_{y_i}^{y_f}dy \int_{z_i}^{1-y}dz\,\left(\frac{z}{y^2}+\frac{y}{z^2} \right)(1-y-z)^3\left[4s-\overline{m}_c^2+\frac{2s^2}{3}\delta \left(s-\overline{m}_c^2\right) \right]  \nonumber\\
&&+\frac{m_c^2}{576\pi^{4}}\langle \frac{\alpha_sGG}{\pi}\rangle\int_{y_i}^{y_f}dy \int_{z_i}^{1-y}dz\,\left(\frac{z}{y^2}+\frac{y}{z^2} \right)(1-y-z)^2\left(3s-2\overline{m}_c^2 \right)   \nonumber\\
&&-\frac{m_c^4}{864\pi^{4}}\langle \frac{\alpha_sGG}{\pi}\rangle\int_{y_i}^{y_f}dy \int_{z_i}^{1-y}dz\,\left(\frac{1}{y^3}+\frac{1}{z^3} \right)(1-y-z)^3 \left[\frac{1}{2}+s\,\delta \left(s-\overline{m}_c^2\right) \right] \nonumber\\
&&+\frac{m_c^2}{576\pi^{4}}\langle \frac{\alpha_sGG}{\pi}\rangle\int_{y_i}^{y_f}dy \int_{z_i}^{1-y}dz\,(1-y-z)\left[1+\left(\frac{1}{y^2}+\frac{1}{z^2} \right)(1-y-z)^2\right] \left(3s-\overline{m}_c^2\right)   \nonumber\\
&&+\frac{m_c^4}{576\pi^{4}}\langle \frac{\alpha_sGG}{\pi}\rangle\int_{y_i}^{y_f}dy \int_{z_i}^{1-y}dz\,\left(\frac{1}{y^3}+\frac{1}{z^3} \right)(1-y-z)^2 \nonumber\\
&&-\frac{m_c^2}{576\pi^{4}}\langle \frac{\alpha_sGG}{\pi}\rangle\int_{y_i}^{y_f}dy \int_{z_i}^{1-y}dz\,\left[\left(\frac{3}{y^2}+\frac{3}{z^2} \right)(1-y-z)^2-4\right] \left(s-\overline{m}_c^2\right)   \nonumber\\
&&+\frac{1}{384\pi^{4}}\langle \frac{\alpha_sGG}{\pi}\rangle\int_{y_i}^{y_f}dy \int_{z_i}^{1-y}dz\, (1-y-z)^2 \left(s-\overline{m}_c^2\right)\left(2s-\overline{m}_c^2\right)    \nonumber\\
&&+\frac{1}{576\pi^{4}}\langle \frac{\alpha_sGG}{\pi}\rangle\int_{y_i}^{y_f}dy \int_{z_i}^{1-y}dz\, yz(1-y-z) \left(25s^2-24s\overline{m}_c^2+3\overline{m}_c^4\right)    \nonumber\\
&&+\frac{1}{576\pi^{4}}\langle \frac{\alpha_sGG}{\pi}\rangle\int_{y_i}^{y_f}dy \int_{z_i}^{1-y}dz\, yz \left(s-\overline{m}_c^2\right)\left(s-2\overline{m}_c^2\right)    \, ,
\end{eqnarray}

\begin{eqnarray}
\rho_{5;1;V}(s)&=&- \frac{m_s\langle\bar{s}g_s\sigma Gs\rangle}{24\pi^{4}}\int_{y_i}^{y_f}dy  \int_{z_i}^{1-y}dz\,yz \left[4s-\overline{m}_c^2+\frac{2s^2}{3}\delta \left(s-\overline{m}_c^2\right) \right]  \nonumber\\
&& +\frac{m_s\langle\bar{q}g_s\sigma Gq\rangle}{48\pi^{4}}\int_{y_i}^{y_f}dy  \, y(1-y) \left(3s-\widetilde{m}_c^2\right)   \nonumber\\
&& -\frac{m_s\left[6\langle\bar{q}g_s\sigma Gq\rangle-\langle\bar{s}g_s\sigma Gs\rangle\right]}{144\pi^{4}}\int_{y_i}^{y_f}dy  \, y(1-y) \left(3s-2\widetilde{m}_c^2\right)   \nonumber\\
&&- \frac{m_s m_c^2\langle\bar{s}g_s\sigma Gs\rangle}{72\pi^{4}}\int_{y_i}^{y_f}dy \int_{z_i}^{1-y}dz \, \left[\frac{1}{2}+s\,\delta \left(s-\overline{m}_c^2\right) \right]   \nonumber\\
&&- \frac{m_s m_c^2\left[9\langle\bar{q}g_s\sigma Gq\rangle-2\langle\bar{s}g_s\sigma Gs\rangle \right]}{144\pi^{4}}\int_{y_i}^{y_f}dy      \nonumber\\
&&+ \frac{m_s\langle\bar{q}g_s\sigma Gq\rangle}{192\pi^{4}}\int_{y_i}^{y_f}dy  \int_{z_i}^{1-y}dz\,(y+z) \left(s-\overline{m}_c^2\right)   \nonumber\\
&&+ \frac{m_s m_c^2\langle\bar{q}g_s\sigma Gq\rangle}{192\pi^{4}}\int_{y_i}^{y_f}dy  \int_{z_i}^{1-y}dz\,\left(\frac{1}{y}+\frac{1}{z}\right)    \, ,
\end{eqnarray}

\begin{eqnarray}
\rho_{6;1;V}(s)&=&-\frac{\langle\bar{q}q\rangle\langle\bar{s}s\rangle }{3\pi^{2}}\int_{y_i}^{y_f}dy \,y(1-y) s \, ,
\end{eqnarray}

\begin{eqnarray}
\rho_{8;1;V}(s)&=&\frac{\langle\bar{s}s\rangle\langle\bar{q}g_s\sigma Gq\rangle+\langle\bar{q}q\rangle\langle\bar{s}g_s\sigma Gs\rangle }{12\pi^{2}}\int_{y_i}^{y_f}dy \,y(1-y)\left[3+\left(2s+\frac{s^2}{T^2}\right)\,\delta\left(s-\widetilde{m}_c^2\right)\right]      \nonumber\\
&&-\frac{\langle\bar{s}s\rangle\langle\bar{q}g_s\sigma Gq\rangle+\langle\bar{q}q\rangle\langle\bar{s}g_s\sigma Gs\rangle }{144\pi^{2}}\int_{y_i}^{y_f}dy \, \left[1+s\,\delta\left(s-\widetilde{m}_c^2\right)\right]\, ,
\end{eqnarray}

\begin{eqnarray}
\rho_{10;1;V}(s)&=& -\frac{ \langle\bar{q}g_s\sigma Gq\rangle  \langle\bar{s}g_s\sigma Gs\rangle }{48\pi^{2}}\int_{y_i}^{y_f}dy \, y(1-y)\,\left(6+\frac{4s}{T^2}+\frac{s^2}{T^4} +\frac{s^3}{T^6}  \right) \delta\left(s-\widetilde{m}_c^2\right)  \nonumber\\
&& +\frac{ \langle\bar{q}g_s\sigma Gq\rangle  \langle\bar{s}g_s\sigma Gs\rangle }{288\pi^{2}}\int_{y_i}^{y_f}dy \, \,\left(2+\frac{s}{T^2}+\frac{s^2}{T^4}   \right) \delta\left(s-\widetilde{m}_c^2\right)  \nonumber\\
&& +\frac{11 \langle\bar{q}g_s\sigma Gq\rangle  \langle\bar{s}g_s\sigma Gs\rangle }{2304\pi^{2}}\int_{y_i}^{y_f}dy \, \,\left(1+\frac{2s}{T^2} \right) \delta\left(s-\widetilde{m}_c^2\right)  \, ,
\end{eqnarray}

\begin{eqnarray}
\rho_{0;2}(s)&=&\frac{1}{960\pi^{6}}\int_{y_i}^{y_f}dy \int_{z_i}^{1-y}dz\,yz(1-y-z)^3\left(s-\overline{m}_c^2\right)^2\left(33s^2-18s\overline{m}_c^2+\overline{m}_c^4\right) \nonumber\\
&&+\frac{1}{480\pi^{6}}\int_{y_i}^{y_f}dy \int_{z_i}^{1-y}dz\,yz(1-y-z)^2\left(s-\overline{m}_c^2\right)^3\left( 9s -2\overline{m}_c^2\right) \nonumber\\
&&+\frac{m_c^2}{288\pi^{6}}\int_{y_i}^{y_f}dy \int_{z_i}^{1-y}dz\,(1-y-z)^3\left(s-\overline{m}_c^2\right)^2 \left( 7s -\overline{m}_c^2\right) \nonumber\\
&&+\frac{m_c^2}{96\pi^{6}}\int_{y_i}^{y_f}dy \int_{z_i}^{1-y}dz\,(1-y-z)^2\left(s-\overline{m}_c^2\right)^3 \, ,
\end{eqnarray}

\begin{eqnarray}
\rho_{3;2}(s)&=&\frac{m_s\langle\bar{s}s\rangle}{60\pi^{4}}\int_{y_i}^{y_f}dy \int_{z_i}^{1-y}dz\,yz (1-y-z)\left(35s^2-30s\overline{m}_c^2+3\overline{m}_c^4\right)  \nonumber\\
&&-\frac{m_s\left[5\langle\bar{q}q\rangle-\langle\bar{s}s\rangle\right]}{60\pi^{4}}\int_{y_i}^{y_f}dy \int_{z_i}^{1-y}dz\,yz \left(s-\overline{m}_c^2\right)\left( 5s -\overline{m}_c^2\right)  \nonumber\\
&&-\frac{m_s\left[10\langle\bar{q}q\rangle-3\langle\bar{s}s\rangle\right]}{60\pi^{4}}\int_{y_i}^{y_f}dy \int_{z_i}^{1-y}dz\,yz \left(s-\overline{m}_c^2\right)\left( 2s -\overline{m}_c^2\right)  \nonumber\\
&&-\frac{m_s m_c^2\left[6\langle\bar{q}q\rangle-\langle\bar{s}s\rangle \right]}{12\pi^{4}}\int_{y_i}^{y_f}dy \int_{z_i}^{1-y}dz\, \left(s-\overline{m}_c^2\right) \nonumber\\
&&+\frac{m_s m_c^2\langle\bar{s}s\rangle }{12\pi^{4}}\int_{y_i}^{y_f}dy \int_{z_i}^{1-y}dz\,(1-y-z) \left(3s-\overline{m}_c^2\right) \, ,
\end{eqnarray}

\begin{eqnarray}
\rho_{4;2}(s)&=&-\frac{m_c^2}{720\pi^{4}}\langle \frac{\alpha_sGG}{\pi}\rangle\int_{y_i}^{y_f}dy \int_{z_i}^{1-y}dz\,\left(\frac{z}{y^2}+\frac{y}{z^2} \right)(1-y-z)^3\left[5s-\overline{m}_c^2+\frac{4s^2}{3}\delta \left(s-\overline{m}_c^2\right) \right]  \nonumber\\
&&-\frac{m_c^2}{1440\pi^{4}}\langle \frac{\alpha_sGG}{\pi}\rangle\int_{y_i}^{y_f}dy \int_{z_i}^{1-y}dz\,\left(\frac{z}{y^2}+\frac{y}{z^2} \right)(1-y-z)^2\left(15s-8\overline{m}_c^2 \right)   \nonumber\\
&&-\frac{m_c^4}{432\pi^{4}}\langle \frac{\alpha_sGG}{\pi}\rangle\int_{y_i}^{y_f}dy \int_{z_i}^{1-y}dz\,\left(\frac{1}{y^3}+\frac{1}{z^3} \right)(1-y-z)^3 \left[\frac{1}{2}+s\,\delta \left(s-\overline{m}_c^2\right) \right] \nonumber\\
&&+\frac{m_c^2}{288\pi^{4}}\langle \frac{\alpha_sGG}{\pi}\rangle\int_{y_i}^{y_f}dy \int_{z_i}^{1-y}dz\,(1-y-z)\left[1+\left(\frac{1}{y^2}+\frac{1}{z^2} \right)(1-y-z)^2\right] \left(3s-\overline{m}_c^2\right)   \nonumber\\
&&-\frac{m_c^4}{288\pi^{4}}\langle \frac{\alpha_sGG}{\pi}\rangle\int_{y_i}^{y_f}dy \int_{z_i}^{1-y}dz\,\left(\frac{1}{y^3}+\frac{1}{z^3} \right)(1-y-z)^2 \nonumber\\
&&+\frac{m_c^2}{288\pi^{4}}\langle \frac{\alpha_sGG}{\pi}\rangle\int_{y_i}^{y_f}dy \int_{z_i}^{1-y}dz\,\left[\left(\frac{3}{y^2}+\frac{3}{z^2} \right)(1-y-z)^2 -5\right] \left(s-\overline{m}_c^2\right)   \nonumber\\
&&-\frac{1}{1440\pi^{4}}\langle \frac{\alpha_sGG}{\pi}\rangle\int_{y_i}^{y_f}dy \int_{z_i}^{1-y}dz\, (1-y-z)^3 \left(15s^2-15s\overline{m}_c^2+2\overline{m}_c^4\right)     \nonumber\\
&&-\frac{1}{960\pi^{4}}\langle \frac{\alpha_sGG}{\pi}\rangle\int_{y_i}^{y_f}dy \int_{z_i}^{1-y}dz\, (1-y-z)^2 \left(s-\overline{m}_c^2\right)\left(3s-2\overline{m}_c^2\right)    \nonumber\\
&&+\frac{1}{1440\pi^{4}}\langle \frac{\alpha_sGG}{\pi}\rangle\int_{y_i}^{y_f}dy \int_{z_i}^{1-y}dz\, yz(1-y-z) \left(35s^2-30s\overline{m}_c^2+3\overline{m}_c^4\right)    \nonumber\\
&&-\frac{1}{1440\pi^{4}}\langle \frac{\alpha_sGG}{\pi}\rangle\int_{y_i}^{y_f}dy \int_{z_i}^{1-y}dz\, yz \left(s-\overline{m}_c^2\right)\left(34s-11\overline{m}_c^2\right)    \, ,
\end{eqnarray}

\begin{eqnarray}
\rho_{5;2}(s)&=&- \frac{m_s\langle\bar{s}g_s\sigma Gs\rangle}{60\pi^{4}}\int_{y_i}^{y_f}dy  \int_{z_i}^{1-y}dz\,yz \left[5s-\overline{m}_c^2+\frac{4s^2}{3}\delta \left(s-\overline{m}_c^2\right) \right]  \nonumber\\
&& +\frac{m_s\left[15\langle\bar{q}g_s\sigma Gq\rangle-2\langle\bar{s}g_s\sigma Gs\rangle\right]}{360\pi^{4}}\int_{y_i}^{y_f}dy  \, y(1-y) \left(3s-\widetilde{m}_c^2\right)   \nonumber\\
&& +\frac{m_s\left[5\langle\bar{q}g_s\sigma Gq\rangle-\langle\bar{s}g_s\sigma Gs\rangle\right]}{120\pi^{4}}\int_{y_i}^{y_f}dy  \, y(1-y) \left(3s-2\widetilde{m}_c^2\right)   \nonumber\\
&&+ \frac{m_s m_c^2\left[9\langle\bar{q}g_s\sigma Gq\rangle-\langle\bar{s}g_s\sigma Gs\rangle \right]}{72\pi^{4}}\int_{y_i}^{y_f}dy      \nonumber\\
&&- \frac{m_s m_c^2\langle\bar{s}g_s\sigma Gs\rangle}{36\pi^{4}}\int_{y_i}^{y_f}dy \int_{z_i}^{1-y}dz \, \left[\frac{1}{2}+s\,\delta \left(s-\overline{m}_c^2\right) \right]   \nonumber\\
&&+ \frac{m_s\langle\bar{q}g_s\sigma Gq\rangle}{96\pi^{4}}\int_{y_i}^{y_f}dy  \int_{z_i}^{1-y}dz\,(y+z) \left(2s-\overline{m}_c^2\right)   \nonumber\\
&&+ \frac{m_s m_c^2\langle\bar{q}g_s\sigma Gq\rangle}{96\pi^{4}}\int_{y_i}^{y_f}dy  \int_{z_i}^{1-y}dz\,\left(\frac{1}{y}+\frac{1}{z}\right)    \, ,
\end{eqnarray}

\begin{eqnarray}
\rho_{6;2}(s)&=&\frac{4\langle\bar{q}q\rangle\langle\bar{s}s\rangle }{3\pi^{2}}\int_{y_i}^{y_f}dy \,y(1-y)s\, ,
\end{eqnarray}

\begin{eqnarray}
\rho_{8;2}(s)&=&-\frac{\langle\bar{s}s\rangle\langle\bar{q}g_s\sigma Gq\rangle+\langle\bar{q}q\rangle\langle\bar{s}g_s\sigma Gs\rangle }{6\pi^{2}}\int_{y_i}^{y_f}dy \,y(1-y)\left[3+\left(4s+\frac{2s^2}{T^2} \right)\,\delta\left(s-\widetilde{m}_c^2\right)\right]      \nonumber\\
&&-\frac{\langle\bar{s}s\rangle\langle\bar{q}g_s\sigma Gq\rangle+\langle\bar{q}q\rangle\langle\bar{s}g_s\sigma Gs\rangle }{72\pi^{2}}\int_{y_i}^{y_f}dy \,\left[1+2s\,\delta\left(s-\widetilde{m}_c^2\right)\right] \, ,
\end{eqnarray}

\begin{eqnarray}
\rho_{10;2}(s)&=& \frac{ \langle\bar{q}g_s\sigma Gq\rangle  \langle\bar{s}g_s\sigma Gs\rangle }{12\pi^{2}}\int_{y_i}^{y_f}dy \, y(1-y)\,\left(\frac{s}{T^2}+\frac{s^2}{T^4} +\frac{s^3}{T^6} \right) \delta\left(s-\widetilde{m}_c^2\right)  \nonumber\\
&& +\frac{ \langle\bar{q}g_s\sigma Gq\rangle  \langle\bar{s}g_s\sigma Gs\rangle }{144\pi^{2}}\int_{y_i}^{y_f}dy \, \,\left(\frac{s}{T^2} +\frac{2s^2}{T^4}  \right) \delta\left(s-\widetilde{m}_c^2\right)  \nonumber\\
&& +\frac{11 \langle\bar{q}g_s\sigma Gq\rangle  \langle\bar{s}g_s\sigma Gs\rangle }{1152\pi^{2}}\int_{y_i}^{y_f}dy \, \,\frac{s}{T^2} \delta\left(s-\widetilde{m}_c^2\right)  \, ,
\end{eqnarray}
  $y_{f}=\frac{1+\sqrt{1-4m_c^2/s}}{2}$,
$y_{i}=\frac{1-\sqrt{1-4m_c^2/s}}{2}$, $z_{i}=\frac{y
m_c^2}{y s -m_c^2}$, $\overline{m}_c^2=\frac{(y+z)m_c^2}{yz}$,
$ \widetilde{m}_c^2=\frac{m_c^2}{y(1-y)}$, $\int_{y_i}^{y_f}dy \to \int_{0}^{1}dy$, $\int_{z_i}^{1-y}dz \to \int_{0}^{1-y}dz$ when the $\delta$ functions $\delta\left(s-\overline{m}_c^2\right)$ and $\delta\left(s-\widetilde{m}_c^2\right)$ appear.

\section*{Acknowledgements}
This  work is supported by National Natural Science Foundation, Grant Number 11375063.

\end{document}